%Paper: 9202007
%From: james@beauty.tn.cornell.edu (James Grochocinski)
%Date: Mon, 3 Feb 92 15:56:56 EST

\input phyzzx
  \tolerance=1000
  \newdimen\fullhsize
    \fullhsize=10truein
    \def\fulline{\hbox to \fullhsize}
  \def\doubleformat{\shipout\vbox{\makeheadline
     \fulline{\box\leftcolumn\hfil\columnbox} }}
  \def\columnbox{\leftline{\vbox{{\pagebody\makefootline}}}}
  \newbox\leftcolumn
  \let\lr=L
\message{Two pages per side (2) or one (1)?}
%%%XXX\read-1 to \twopage
\def\twopage{1}
\if 2\twopage
  \message{***Print this file in landscape mode***}

  \baselineskip=16pt plus 2pt minus 1pt
  \hsize=4.75truein           \vsize=7truein
  \hfuzz=5pt
  \hoffset=-.56truein	\voffset=-.4truein
  \output={\if L\lr
    \global\setbox\leftcolumn=\columnbox \global\let\lr=R
    \else\doubleformat \global\let\lr=L\fi
    \advancepageno
    \ifnum\outputpenalty>-20000 \else\dosupereject\fi}
  \def\shrinker{\scriptstyle}
\else
  \def\shrinker{\relax}
\fi

\overfullrule=0pt

\def\bZ{{\bf Z}}
\def\ba{\langle}
\def\fa{\rangle}
\def\pv{\partial\varphi}
\def\pvp#1{\partial^#1\varphi}
\def\B#1{\beta_{#1}}           %\B{12}
\def\la#1{\lambda_{#1}}        %\la{12}
\def\lb#1{\bar\la{#1}}        %\lb{12}
\def\tci{${\rm TCI}^2$}
\def\ph#1#2#3{\Phi^{(#1)}_{#2,#3}}   %\ph1 1 2
\def\lph#1#2#3#4{(L_{-#4}\Phi)^{(#1)}_{#2,#3}}    %\lph1 2 2 3
\def\fpi#1#2{\Phi_{#1,#2}}
\def\w#1{\omega_#1}
\def\sf#1#2{\sqrt{#1\over#2}}
\def\simdot{\buildrel . \over \sim}
\def\h#1{\hat#1}
\def\mtt#1#2#3#4{\pmatrix{#1&#2\cr#3&#4\cr}}
\def\vr#1{\{#1\}}
\def\vrv#1{\{{\vec#1}\}}
\def\tj#1{$\{T,J^{(#1)}\}$}
\def\hs#1{H_#1}
\def\hh{$\{T,J,H_\ell\}$}
\def\ff{{\cal F}}
\def\pp{{\cal P}}
\def\zk#1{$Z_{#1}$ PF}
\def\as#1{\alpha_#1}
\def\ie{{\it i.e.\/}}
\def\eg{{\it e.g.\/}}
\def\p#1{\Phi_{#1,1}}
\def\f#1#2#3{f_{#1#2#3}}
\def\srx{\sqrt{rx}}
\def\fb{\bar5}
\def\fr#1#2{{#1\over#2}}
\def\kl#1#2{$[\![#1,#2]\!]$}

\Pubnum={CLNS 91/1126}
\date{January 1992}
\pubtype={ }
\titlepage
\bigskip
\title{\bf Construction of the K=8 Fractional Superconformal Algebras}
\author{Philip C. Argyres\foot{pca@strange.tn.cornell.edu,
pca@crnlnuc.bitnet},
James M. Grochocinski\foot{james@beauty.tn.cornell.edu,
james@crnlnuc.bitnet},
and S.-H. Henry Tye}
\address{Newman Laboratory of Nuclear Studies \break
Cornell University \break
Ithaca, NY 14853-5001}
\abstract
{We construct the $K=8$ fractional superconformal algebras.
There are two such extended Virasoro algebras, one of which
was constructed earlier, involving a fractional spin
(equivalently, conformal dimension) ${6\over5}$ current.
The new algebra involves two additional fractional spin
currents with spin ${13\over5}$.  Both algebras are non-local
and satisfy non-abelian braiding relations.  The construction
of the algebras uses the isomorphism between the $Z_8$
parafermion theory and the tensor product of two tricritical
Ising models.  For the special value of the central charge
$c={52\over55}$, corresponding to the eighth member of the
unitary minimal series, the ${13\over5}$ currents of the new
algebra decouple, while two spin ${23\over5}$ currents (level-$2$
current algebra descendants of the ${13\over5}$ currents)
emerge. In addition, it is shown that the $K=8$ algebra
involving the spin ${13\over5}$ currents at central charge
$c={12\over5}$ is the appropriate algebra for the construction
of the $K=8$ (four-dimensional) fractional superstring.}
\endpage

%1
 \REF\BPZ{A.A.\ Belavin, A.M.\ Polyakov and A.B.\ Zamolodchikov,
 {\sl Nucl.\ Phys.} {\bf B241} (1984) 333.}
%2
 \REF\FQS{D.\ Friedan, Z.\ Qiu and S.\ Shenker, {\sl Phys.\ Rev.\ Lett.}
 {\bf 52} (1984) 1575.}
%3
 \REF\Knizh{V.G.\ Knizhnik and A.B.\ Zamolodchikov, {\sl Nucl.\ Phys.}
 {\bf B247} (1984) 83.}
%4
 \REF\Zam{A.B.\ Zamolodchikov, {\sl Theor.\ Math.\ Phys.} {\bf 63}
 (1985) 1205.}
%5
 \REF\ACT{C.\ Ahn, S.-w.\ Chung and S.-H.H.\ Tye, {\sl Nucl.\ Phys.}
 {\bf B365} (1991) 191.}
%6
 \REF\Freidan{See \eg\ D.\ Freidan in {\it Unified String Theories,}
 ed.\ by M.\ Green and D.\ Gross, World Scientific (1986).}
%7
 \REF\ZaFa{A.B.\ Zamolodchikov and V.A.\ Fateev,
 {\sl Sov.\ Phys.\ J.E.T.P.} {\bf 62} (1985) 215.}
%8
 \REF\agt{P.C.\ Argyres, J.\ Grochocinski and S.-H.H.\ Tye,
 {\sl Nucl.\ Phys.} {\bf B367} (1991) 217.}
%9
 \REF\GKO{P.\ Goddard, A.\ Kent and D.\ Olive, {\sl Comm.\ Math.\ Phys.}
 {\bf 103} (1986) 105.}
%10
 \REF\ZFft{V.A.\ Fateev and A.B.\ Zamolodchikov, {\sl Sov.\ Phys.\ J.E.T.P.}
 {\bf 71} (1988) 451.}
%11
 \REF\Kastor{D.\ Kastor, E.\ Martinec and Z.\ Qiu, {\sl Phys.\ Lett.}
 {\bf 200B} (1988) 434;  J.\ Bagger, D.\ Nemeschansky and
 S.\ Yankielowicz, {\sl Phys.\ Rev.\ Lett.} {\bf 60} (1988) 389;
 F.\ Ravanini, {\sl Mod.\ Phys.\ Lett.} {\bf A3} (1988) 397.}
%12
 \REF\CLT{S.-w.\ Chung, E.\ Lyman and S.-H.H.\ Tye,
 {\sl Int.\ J.\ Mod.\ Phys.} {\bf A7} \#14 (1992).}
%13
 \REF\DoFa {Vl.S.\ Dotsenko and V.A.\ Fateev, {\sl Nucl.\ Phys.}
 {\bf B240} (1984) 312; {\sl Nucl.\ Phys.} {\bf B251} (1985) 691.}
%14
 \REF\Dots{Vl.S.\ Dotsenko, {\sl Nucl.\ Phys.} {\bf B338} (1990) 747;
 {\sl Nucl.\ Phys.} {\bf B358} (1991) 547.}
%15
 \REF\Qiu{Z.\ Qiu, {\sl Nucl.\ Phys.} {\bf B270} (1986) 205; D.\ Friedan,
 Z.\ Qiu and S.\ Shenker, {\sl Phys.\ Lett.} {\bf 151B} (1985) 37.}
%16
 \REF\FZWZW{A.B.\ Zamolodchikov and V.A.\ Fateev, {\sl Sov.\ J.\
 Nucl.\ Phys.} {\bf 43} (1986) 657.}
%17
 \REF\Felder{G.\ Felder, {\sl Nucl.\ Phys.} {\bf B317} (1989) 215.}
%18
 \REF\ArT{P.C.\ Argyres and S.-H.H.\ Tye, {\sl Phys.\ Rev.\ Lett.}
 {\bf 67} (1991) 3339.}
%19
 \REF\Rocha{A.\ Rocha-Caridi in {\it Vertex Operators in Mathematics and
 Physics,} MSRI Publication {\bf 3} (Springer, Heidelberg 1984) 451.}
%20
 \REF\Kac{V.G.\ Kac and D.\ Peterson, {\sl Bull.\ AMS} {\bf 3}
 (1980) 1057; {\sl Adv.\ Math.} {\bf 53} (1984) 125;
 J.\ Distler and Z.\ Qiu, {\sl Nucl.\ Phys.} {\bf B336} (1990) 533.}
%21
 \REF\FSS{K.R.\ Dienes and S.-H.H.\ Tye, {\it Model-Building for Fractional
 Superstrings,} Cornell preprint CLNS 91/1100, McGill preprint McGill/91-29
 (November 1991); P.C.\ Argyres, K.R.\ Dienes and S.-H.H.\ Tye, {\it New
 Jacobi-like Identities for $Z_K$ Parafermion Characters,} CLNS 91/1113,
 McGill/91-37 (January 1992); P.C.\ Argyres, E.\ Lyman and S.-H.H.\ Tye,
 {\it {Low-Lying States of the Six-Dimensional Fractional Superstring,}}
 CLNS 91/1121 (February 1992).}

\chapter{Introduction}

The structure of two-dimensional conformal field theories (CFT) is in
large part determined by their underlying conformal symmetry algebra
[\BPZ].  This infinite dimensional algebra organizes all the fields in a
CFT into sets of primary fields of definite conformal dimensions,
and their associated infinite towers of descendant fields.  The
fundamental conformal symmetry is the Virasoro algebra,
 $$
  T(z)T(w) = {{c/2}\over{(z-w)^4}} + {{2T(w)}\over{(z-w)^2}}
  + {{\partial T(w)}\over{(z-w)}} + \cdots
  \eqn\Ii
 $$
where $T(z)$ is the energy-momentum tensor and the ellipsis
refers to further Virasoro descendants of the identity.  It turns
out that for $c<1$ this symmetry actually determines all unitary
models and their complete spectrum [\BPZ,\FQS].

It is natural to try to classify $c\geq1$ CFTs by
extending the conformal symmetry with new currents [\Knizh,\Zam].
The most general extended conformal symmetries can be written
as follows. Consider a set of currents $J_i(z)$, primary with respect
to $T(z)$, with conformal dimensions $h_i$, where $i\in\{1,2,\ldots,N\}$.
In other words
 $$
  T(z)J_i(w) = {{h_iJ_i(w)}\over{(z-w)^2}}
  +{{\partial J_i(w)}\over{(z-w)}}+\cdots
  \eqn\Iii
 $$
The operator product expansions (OPEs) among the
currents have the generic form
 $$\eqalign{
  J_i(z)J_j(w)=&~q_{ij}(z-w)^{-h_i-h_j}(1+\cdots)\cr
  &~~~+\sum_kf_{ijk}(z-w)^{-h_i-h_j+h_k}
  \left(J_k(w)+\cdots\right),\cr}
  \eqn\Iiii
 $$
where $q_{ij}$ and $f_{ijk}$ are structure constants. The ellipses
stand for current algebra descendants,
whose dimensions differ {}from those of
the identity and the $J_k$ by positive integers.
Since the algebra is chiral,
\ie\ independent of $\bar z$, the conformal dimensions are the spins of
the fields. The parameters $c$, $h_i$, $q_{ij}$
and $f_{ijk}$ are not free and
must be chosen such that the algebra \Iiii\ is associative.  This condition
places strong constraints on the set of consistent conformal dimensions
$h_i$ and restricts the structure constants $q_{ij}$ and $f_{ijk}$ as
functions of the central charge $c$.  We note that there are known examples
where the central charge itself is restricted (\eg\ the spin 5/2 current
algebra of ref.~[\Zam]), and there are also known examples where the
number of currents, $N$, is infinite (\eg\ the $Z_{t/u}$ parafermion theory
of ref.~[\ACT]).

Extended conformal algebras naturally fall into three classes:

(1) Local algebras: the simplest type, where all powers of $(z-w)$
appearing in \Iiii\ are integers. This class includes the most familiar
extended algebras, such as the superconformal, Ka\v{c}-Moody and $W_n$
algebras.  These examples are unitary and therefore consist of currents
with only integer and half-integer spins.  Additionally, there exist
non-unitary algebras such as ghost systems where arbitrary spins may
be present [\Freidan].

When fractional powers of $(z-w)$ appear in \Iiii, some of the currents
will necessarily have fractional spins. In this case, the algebra is
non-local, due to the presence of Riemann cuts in the complex plane.
Such algebras are more complicated to construct and analyze than the
local ones.  Among non-local algebras there is a further division, again
along lines of complication.

(2) Abelian non-local algebras: also known as parafermion (PF)
or generalized parafermion current algebras, were first constructed by
Zamolodchikov and Fateev [\ZaFa].  They are the simplest type of
non-local algebras, involving at most one fractional power of $(z-w)$
in each OPE in \Iiii.  Any two currents
in a PF algebra obey abelian braiding
relations, \ie\ upon braiding teh two currents
(analytically continuing one current along a path
encircling the other), any correlation function involving these two
currents only changes by a phase.  The analysis of the associativity
conditions for PF theories can be carried out using algebraic methods.

(3) Non-abelian non-local algebras: or non-abelian algebras for short,
since they are necessarily non-local. This is the most general class of
extended algebras and is the focus of this paper.  Their characteristic
feature is that their OPEs involve multiple cuts, \ie\ there are terms in at
least one of the OPEs in \Iiii\ with different fractional powers of $(z-w)$.
Any two fractional spin currents appearing in one of these OPEs will in
general obey non-abelian braiding properties.  The analysis of the
associativity conditions for non-abelian algebras requires more powerful
methods.  The first set of non-abelian algebras were constructed in
ref.~[\agt].

In general, the holomorphic $n$-point correlation functions of the
currents,
 $$
 \ba J_i(z_1)J_j(z_2)\cdots J_k(z_n)\fa~,
 \eqn\Iiiia
 $$
can be expressed
as a linear combination of some set of conformal blocks.  The relative
coefficients of the various conformal blocks are fixed by the closure
condition and the associativity condition.  The closure condition
is simply the requirement that no new currents beyond the currents of
the algebra should appear.  Associativity is the condition that the
particular linear combination of conformal blocks that appears in the
$n$-point function is invariant under fusion transformations (\ie\
duality).  For the local algebras, each conformal block involves only
integer powers of $(z_i-z_j)$; for the abelian non-local algebras, each
correlation function of the parafermion currents has exactly one
conformal block, even though it involves fractional powers of $(z_i-z_j)$.
Of course, for the most general case we expect each correlation function
to have multiple conformal blocks, and the conformal blocks to involve
different fractional powers of $(z_i-z_j)$.  This general case corresponds
to the non-abelian non-local algebras.  {}From this point of view we see
that upon braiding the currents, the correlation function \Iiiia\ is, in
general, transformed into an independent linear combination of conformal
blocks.  This reflects the different phases that are picked up upon
analytically continuing the different fractional powers of $(z_i-z_j)$.

The different braiding properties of the different types of extended
algebras described above are reflected in the moding of their currents.
On a given state in any representation of the algebra, we can obtain new
states by acting with current modes
 $$
  J^{i_1}_{-n_1-r_1}J^{i_2}_{-n_2-r_2}\cdots J^{i_m}_{-n_m-r_m}|\Phi\fa,
  \eqn\Iiv
 $$
where the $n_j$ are integers and the $r_j$ are fractional in general.  For
local unitary algebras, the half-integer spin currents can have only integer
or half-integer modings, and $r_1=r_2=\cdots=r_m=r$ where either $r=0$
or $r={1\over2}$, depending on the state $|\Phi\fa$. For PF theories, the
situation is slightly more complicated [\ZaFa].  Generically the $r_i$ are
different, determined by the state $|\Phi\fa$ and the currents that
preceded it. For non-abelian algebras, the moding of a particular current
in \Iiv\ is not unique; it depends both on the state it operates on as
well as on the state we want it to create.

Now, the existence of non-abelian algebras and their usefulness in
organizing CFTs is illustrated by
the $SU(2)$ WZW coset models. Let us denote the
$SU(2)_K\otimes SU(2)_L/SU(2)_{K+L}$ coset model by \kl KL. It is
well known that the \kl1L coset series are exactly the unitary
models with $c=1-{6\over(L+2)(L+3)}$ [\GKO], so that they
are representations of the Virasoro algebra.
Next, the \kl2L coset series are
representations of the superconformal algebra $\{T(z),J^{(2)}(z)\}$,
where $J^{(2)}(z)$ is the usual supercurrent. It is also
known that the \kl4L coset series are representations of the
parafermion current algebra $\{T(z),\psi_1(z),\psi_2(z)\}$, or
$\{T(z),J^{(4)}(z)\}$ where the spin ${4\over3}$ current is
$J^{(4)}(z)=\psi_1(z)+\psi_2(z)$ [\ZFft]. This pattern strongly suggests the
existence of extended algebras for other values of $K$. This belief
was further supported by an explicit construction of the branching
functions of the \kl KL coset series based on the assumption that
an extended Virasoro symmetry exists [\Kastor].
It turns out that the extended
Virasoro algebras for $K$ other than 2 and 4 are not parafermionic,
hence they must be non-abelian.

Some of the tools needed for the analysis and construction of special
series of non-abelian algebras have already been developed.  In particular,
in ref.~[\agt] we constructed a series of non-abelian algebras, the
fractional superconformal algebras (FSCAs), so-called because they
generalize the conventional superconformal algebras. The FSCAs
constructed in ref.~[\agt] are minimal in the sense that they contain only
one fractional spin current, $J^{(K)}(z)$, in addition to the
energy-momentum tensor $T(z)$.  This \tj K algebra
describes the simplest extended conformal symmetry underlying the
\kl KL coset models [\Kastor,\CLT].
The explicit form of this algebra is
 $$\eqalign{
  T(z)T(w)=&~{c\over 2}(z-w)^{-4}\left\{1+\cdots\right\},\cr
  T(z)J^{(K)}(w)=&~\Delta (z-w)^{-2}\left\{J^{(K)}(w)+\cdots\right\},\cr
  J^{(K)}(z)J^{(K)}(w)=&~(z-w)^{-2\Delta}\left\{1+\cdots\right\}\cr
  &~~~~+\lambda_K(c)(z-w)^{-\Delta}
  \left\{J^{(K)}(w)+\cdots\right\},\cr}
  \eqn\Iv
 $$
where $\Delta=(K+4)/(K+2)$ is the dimension of the $J^{(K)}$ current.
The structure constant $\lambda_K(c)$ has the following form [\agt]
 $$
  \lambda^2_K (c) = {{2 K^2 (c_{111})^2}\over{3(K+4)(K+2)}}
  \left[{{3 (K+4)^2}\over{K(K+2)}}{1\over c} - 1\right],
  \eqn\Ivi
 $$
where
 $$
  (c_{111})^2={{\sin^2(\pi\rho)\sin^2(4\pi\rho)}
  \over{\sin^3(2\pi\rho)\sin(3\pi\rho)}}
  {{\Gamma^3(\rho)\Gamma^2(4\rho)}
  \over{\Gamma(3\rho)\Gamma^4(2\rho)}},
  \eqn\Ivii
 $$
and $\rho={1\over{K+2}}$. The structure constant $c_{111}$ is that for
the OPE of two spin-one fields to close on another spin-one field in the
chiral $SU(2)_K$ WZW model where all other higher-spin fields are
decoupled.

In this paper we shall consider a more complex example of a non-abelian
extended algebra which has three fractional spin currents.  Specifically,
we will show that there are two consistent $K=8$ FSCAs.  One is the
algebra given in \Iv\ for $K=8$.  The other is a new algebra which
involves,  besides the spin $6\over5$ current $J$, two additional
currents $\hs1$ and $\hs2$, both with spin ${{13}\over5}$.  The form
for this $\{T(z),J(z),\hs1(z),\hs2(z)\}$ algebra will be shown to be
 $$\eqalign{
  J(z)J(w)=&~(z-w)^{-{{12}\over5}}\{1+\cdots\}\cr
  &~\ +s^2\Lambda(c)(z-w)^{-{6\over5}}\{J(w)+\cdots\}\cr
   &~\ \ +s\Omega(c)(z-w)^{1\over5}
    \{\hs1(w)+\cdots+\hs2(w)+\cdots\},\cr
  &\cr
  J(z)\hs1(w)=&~s\Omega(c)(z-w)^{-{{13}\over5}}\{J(w)+\cdots\}\cr
   &~\ \ +{{13}\over{14}}\Lambda(c)
    (z-w)^{-{6\over5}}\{\hs2(w)+\cdots\},\cr
  &\cr
  \hs1(z)\hs1(w)=&~(z-w)^{-{{26}\over5}}\{1+\cdots\}\cr
   &~\ \ -s\Upsilon(c)(z-w)^{-{{13}\over5}}\{\hs1(w)+\cdots\},\cr
  &\cr
  \hs1(z)\hs2(w)=&~{{13}\over{14}}\Lambda(c)
   (z-w)^{-4}\{J(w)+\cdots\},\cr}
  \eqn\Iviii
 $$
where the structure constants are given by
 $$\eqalign{
  &s=\sqrt{{2\over3}rx}~, \qquad r={{\sqrt5-1}\over2}~, \qquad
  x={{\Gamma^2\left({2\over5}\right)}\over
  {\Gamma\left({1\over5}\right)\Gamma\left({3\over5}\right)}}~,\cr
  &\quad\Lambda(c)=\sf{8(27-5c)}{25c}~,\qquad
  \Omega(c)=\sf{18(55c-52)}{455c}~,\cr}
  \eqn\Iix
 $$
and
 $$
  \Upsilon(c)={{\sqrt8(865c-1976)}\over{5\sqrt{455c(55c-52)}}}~.
  \eqn\Ix
 $$
The OPEs involving $T(z)$ follow easily {}from \Ii\ and \Iii, and the
remaining $\hs2(z)$ OPEs are found by exchanging $\hs1\leftrightarrow\hs2$
in \Iviii.

It is important to point out that the \tj8 FSCA is {\it not} a subalgebra of
the \hh\ FSCA, where $\ell=1,2$. They do, however, both have the
Virasoro subalgebra in
common, and in addition $\{T,\hs1\}$ and $\{T,\hs2\}$ are subalgebras of \hh.
For the purposes of classifying CFTs, the existence of two distinct $K=8$
FSCAs shows that there are two independent symmetries that can be used
in the construction of the \kl8L coset
models.  The differences between the representation theories of these
two FSCAs will be discussed in Sect.~7.

As was mentioned above, the \kl KL coset model has an underlying
fractional superconformal symmetry generated by the FSCA. A FSCA current
operating on the identity generates a state with precisely its conformal
dimension, so that the Virasoro primary field in the appropriate
\kl KL coset model with that same conformal dimension can be
identified as the FSCA current. In particular,
all \kl8L coset models, with $L\ge2$, have a ${13\over5}$ primary field
to associate with the $\hs\ell$ FSCA currents. However, the \kl18
coset model, corresponding to the eighth member of the unitary series,
has no such dimension ${13\over5}$ primary field. This apparent
inconsistency is resolved when we note that the central charge of this
model is $c={52\over55}$. At precisely this value of $c$, the
$\hs\ell$ currents become null and decouple {}from the algebra \Iviii.
Instead, in the $c={52\over55}$ unitary
model we find a dimension ${23\over5}$ primary field. For
$c>{52\over55}$ there also exist ${23\over5}$ currents, denoted by
$\hs1'$ and $\hs2'$, but they can be considered as level-2 current algebra
descendants of $\hs1$ and $\hs2$. For $c={52\over55}$, however, since
the $\hs\ell$ are null, the $\hs\ell'$ are promoted to be
FSCA primary currents. In Sect.~7, calculating
directly in the $c={52\over55}$
unitary model using the methods of Dotsenko and Fateev [\DoFa], we
compute the structure constants for this $\{T,J,\hs\ell'\}$ FSCA.
The agreement we find for the $\ba JJJ\fa$ structure constant between
this calculation and the result of eqs. \Iviii\ and \Iix\ provides
a non-trivial check on the calculations carried out in this paper.

Although the main results \Iv-\Ix\ are representation independent, in
this paper we will construct the \tj8 and \hh\ FSCAs by using a special
representation of the \zk8\ theory. The connection between the $K=8$
FSCAs and the \zk8\ theory arises as follows.  Since the \kl8L
coset models form representations of the $K=8$
FSCAs, so does the $SU(2)_8$ chiral WZW model, since it is the
$L\rightarrow\infty$ limit of the coset models.   In particular, at the
special value of the central charge $c={12\over5}$ (the $SU(2)_8$ central
charge), we expect the $K=8$ FSCA to be embedded in the operator algebra
of the $SU(2)_8$ WZW model.  Thus, to construct the $K=8$ FSCAs
for this special value of the central charge, we need to identify $SU(2)_8$
Virasoro primary fields of the appropriate dimensions with the currents
$J$, $\hs1$ and $\hs2$, and then solve the associativity constraints for the
chiral $SU(2)_8$ structure constants in order to calculate the $J$, $\hs1$
and $\hs2$ OPEs.  Unfortunately, there are over a hundred potentially
non-trivial structure constants in the $SU(2)_8$ WZW model, and among
them a great many more associativity constraints (which, though not all
independent, must all be checked).  Although one could in principle perform
a direct computation in the $SU(2)_8$ theory [\Dots], we use instead the
following procedure which makes this problem more tractable.

Begin by representing, in the standard way, the $SU(2)_8$ WZW model
as the tensor product of a \zk8 and a free boson $\varphi(z)$.
In Sect.~2 we show that the \zk8 model (which has central charge
$c={7\over5}$) is isomorphic to the tensor product of two
tricritical Ising (TCI) models [\FQS,\Qiu] (each of which has
$c={7\over10}$).  The usefulness of this
observation resides in the fact that the TCI model has only five primary
fields (besides the identity) and a manageable number of associativity
constraints relating the structure constants of their OPEs.  In Sect.~3 we
solve these associativity conditions for the chiral TCI model and in
Sect.~4 we show how all the associative solutions to the \tci\ model
can be constructed.  The TCI primary fields play the role of a useful
book-keeping device for organizing the \zk8 (or \tci) fields.   In
particular we show that there are only four
inequivalent solutions of the \tci\ associativity constraints, called
$\ff1$, $\ff2$, $\pp1$ and $\pp2$, two of which ($\ff1$ and $\pp1$)
form the basis for
constructing the \hh\ FSCA, while the other two are related to the \tj8
algebra in the same way.  (We are not counting as distinct algebras the
many associative solutions which can be obtained as subalgebras of these
four solutions simply by decoupling various sets of fields.)  We also give a
set of simple rules for calculating the structure constants of these
operator product algebras starting {}from the structure constants of the
chiral TCI model found in Sect.~3.

Given the structure constants of the \zk8 theory (\ie\ the \tci\ model)
found in Sect.~4, we can add back in the free boson $\varphi(z)$, implicitly
forming associative solutions for the chiral $SU(2)_8$ WZW model.  We
can then identify the FSCA currents and calculate their OPEs using our
knowledge of the \zk8 structure constants, and so derive the
algebras \Iv-\Ix\ for the special value of the central charge
$c={12\over5}$.  In particular, in Sect.~5 we derive the \hh\ FSCA for
$c={12\over5}$ in this way {}from the $\ff1$ (or $\pp1$) \tci\ model by
demanding that the OPEs of those currents close among themselves.
To carry this calculation through in terms of the \tci\ fields and the
free boson $\varphi(z)$, we find we must calculate the form and
normalizations of many Virasoro descendant fields using the conformal
Ward identities following {}from the Virasoro algebra \Ii.  This is a
second advantage of using the \tci\ isomorphism over calculating
directly in the chiral $SU(2)_8$ WZW theory, because the Ka\v{c}-Moody
current algebra Ward identities [\FZWZW] are substantially more difficult
to use.

We construct the \hh\ FSCA for arbitrary central charge in Sect.~6 by
turning on a background charge for $\varphi$ and demanding closure
of the operator product algebra.  In addition, using the $\ff1$ \tci\
structure constants, we construct appropriate screening charges and show
that they commute with all the currents in the \hh\ FSCA with background
charge.  These can then be used to solve for the spectrum and correlation
functions of the FSCA using Feigin-Fuchs techniques [\DoFa,\Felder,\CLT].
In Sect.~7 we carry out the same steps starting with the $\ff2$ (or $\pp2$)
\tci\ model and a boson with background charge
to construct the \tj8 algebra, recovering the results \Iv-\Ivii\ of
ref.~[\agt].  Using the $\ff2$ operator algebra we can also construct the
relevant screening charges for the the \tj8 FSCA.  At this point we
will make a few comments on the differences between the representation
theories of the two $K=8$ FSCAs we have constructed. At the central
charge $c={52\over55}$, the $\hs\ell$ decouple and the FSCA changes {}from
\hh\ to $\{T,J,\hs\ell'\}$. We use the $c={52\over55}$ unitary model
to construct this latter FSCA and then we compare it with the \hh\ and
\tj8 FSCAs.

The motivation for studying the $K=8$ FSCAs in particular (out of all
possible $K$) is the observation that this algebra, at central charge
$c={12\over5}$ (\ie\ zero background charge), is the world-sheet basis
of the $K=8$ fractional superstring theory with critical
space-time dimension four [\ArT].  In Sect.~8 we show that the
single-current algebra \tj8 does not allow the coupling of space-time
fermions in the fractional superstring, whereas the \hh\ algebra does.
Indeed, we show explicitly how to derive the space-time Dirac equation
satisfied by the massless fermion states of the fractional superstring
using the \hh\ algebra.

Finally, we have collected technical discussions in three appendices.
In Appendix A we derive the associativity conditions for the chiral TCI
model following Dotsenko and Fateev [\DoFa].  Appendix B is a compilation
of a special subset of the associative structure constants for the $\ff1$
\tci\ model found in Sect.~4.  In Appendix C we compute the OPEs between
various Virasoro descendant fields in terms of the structure
constants for the primary field OPEs using the conformal Ward identities.

\chapter{The $Z_8$-parafermion and the tricritical Ising model}

After a brief introduction to the \zk8 theory we show how it
corresponds to the tensor product of two TCI models.

The operator content of the chiral \zk8 theory can be realized by the
$SU(2)_8/U(1)$ coset model [\ZaFa].  The chiral $SU(2)_8$ WZW theory
[\Knizh] has central charge $c_{\rm WZW}={12\over5}$ and consists of
holomorphic primary fields $\Phi^j_m(z)$ of conformal dimension
$j(j+1)/10$.  The indices $j,m~\in{\bf Z}/2$ label $SU(2)$ representations
where $0\leq j\leq 4$ and $|m|\leq j$ with $j-m\in \bZ$.  When we factor a
$U(1)$ subgroup out of $SU(2)_8$, we correspondingly factor the primary
fields as
 $$
 \Phi^j_m(z) ~=~ \phi^j_m(z) \,
 \exp\left\{{m\over2}\,\varphi(z) \right\}~.
 \eqn\IIi
  $$
Here $\varphi$ is the free $U(1)$ boson normalized
so that $\langle\varphi(z)\varphi(w)\rangle=+{\rm ln}(z-w)$.
The $\phi^j_m(z)$
are Virasoro primary fields in the \zk8 theory with conformal dimensions:
 $$
 h^j_m~=~{j(j+1)\over10}~-~{m^2\over8}
 \quad\quad\quad{\rm for}~~|m|\leq j~.
 \eqn\IIii
 $$
The central charge of the \zk8 theory is then $c=c_{\rm WZW}-
c_{\varphi}={7\over5}$.  The definition of the $\phi^j_m$ fields can be
consistently extended to the case where $|m|>j$ by the rules
 $$
 \phi^j_m~=~\phi^j_{m+8}~=~\phi^{4-j}_{m-4}~.
 \eqn\IIiii
 $$
The fusion rules of the parafermion fields follow {}from those of the
$SU(2)_8$ theory:
 $$
 [\phi^{j_1}_{m_1}]\,\times\,[\phi^{j_2}_{m_2}]~\sim
 \sum_{j=|j_1-j_2|}^r[\phi^j_{m_1+m_2}]
 \eqn\IIiv
 $$
where $r={\rm min}\{j_1+j_2\, ,\, 8-j_1-j_2\}$. The sectors $[\phi^j_m]$
include the primary fields $\phi^j_m$ and a tower of higher-dimension
fields (with dimensions differing by integers) defined as in \IIi\ {}from
the Ka\v{c}-Moody current algebra descendants of the $\Phi^j_m$.

It follows {}from the fusion rules \IIiv\ that a special set of fields,
called PF currents, form a closed algebra.  The PF currents, denoted
$\psi_\ell$, are defined by
 $$
 \psi_\ell~\equiv~\phi^0_\ell~=~\phi^0_{\ell-8}~.
 \eqn\IIv
 $$
By \IIiii\ and \IIii, the $\psi_\ell$ have conformal dimensions
$\ell(8-\ell)/8$.  The operator product algebra they satisfy is called
the \zk8 current algebra:
 $$
 \psi_\ell(z)\psi_m(w)~=~{c_{\ell,m}\over(z-w)^{s(\ell,m)}}
 \psi_{\ell+m}(w)+\cdots~,
 \eqn\IIvi
 $$
where, if $-4\leq\ell,m\leq4$, then $s(\ell,m)={\ell m\over4}+|\ell|+|m|
-|\ell+m|$.  The currents are normalized so that $c_{\ell,-\ell}=1$.  Notice,
in particular, that the $\psi_4$ PF current has conformal dimension $2$ and
satisfies the OPE
 $$
 \psi_4(z)\psi_4(w)~=~{1\over(z-w)^4}+{(4/c)T_{Z8}(w)\over(z-w)^2}
 +{(2/c)\partial T_{Z8}(w)\over(z-w)}+\cdots
 \eqn\IIvii
 $$
where, for example, the factor $4/c$ in front of the \zk8
energy-momentum tensor $T_{Z8}(w)$ follows {}from
conformal invariance (since $T_{Z8}$ is itself
a descendant of the identity) and
the ellipsis denotes further descendants of
the identity.  Here $c={7\over5}$
is the central charge of the \zk8 theory.

Now we will examine in some depth an interesting and useful
representation of the \zk8 theory.  This representation does not
realize the full set of \zk8 fields, but does realize the subset of
fields (forming a closed operator product algebra) necessary
for the construction of the $K=8$ FSCAs.

Zamolodchikov observed [\Zam] that an operator product algebra with
central charge $c$ and a dimension 2 operator which, together with the
energy-momentum tensor, forms a closed operator subalgebra can be
written in a new basis to be the direct product of two algebras each
with central charge ${c\over2}$.  We noted above that the \zk8 has an
additional dimension $2$ operator besides the energy-momentum tensor
$T_{Z8}(z)$, namely the current $\psi_4(z)$.  Specifically, {}from the
Virasoro algebra \Ii, the $\psi_4\psi_4$ OPE \IIvii\ and the fact that
$\psi_4$ is a primary dimension $2$ field
 $$
  T_{Z8}(z)\psi_4(w) = {{2\psi_4(w)}\over{(z-w)^2}}
  + {{\partial\psi_4(w)}\over{(z-w)}} + \dots~,
  \eqn\IIviii
 $$
it follows that the two combinations
 $$\eqalign{
  T_1=&~{1\over2}\left(T_{Z8}+\sf c2\psi_4\right)~,\cr
  T_2=&~{1\over2}\left(T_{Z8}-\sf c2\psi_4\right)~,\cr}
  \eqn\IIix
 $$
satisfy separate Virasoro algebras with central charge $c/2$, and
that the $T_1T_2$ OPE is regular.  Since the \zk8 theory has central
charge $c={7\over5}$, it follows that $T_1$ and $T_2$ are the
energy-momentum tensors for two (distinct) $c={7\over10}$ CFTs.
Furthermore, since the \zk8 theory is unitary and the only unitary
$c={7\over10}$ conformal field theory is the tricritical Ising (TCI)
model [\FQS], it is
natural to investigate writing the \zk8 theory as the tensor product
of two TCI models (which we will denote by \tci).

The first thing to check is whether or not the fields in the \zk8 model
have corresponding fields in the \tci\ model. The dimensions $\Delta_{r,s}$
of the primary fields $\Phi_{r,s}$ in the TCI model are [\BPZ]
 $$
  \Delta_{r,s}={{(5r-4s)^2-1}\over{80}},
  \quad1\leq r\leq3,\quad1\leq s\leq4.
  \eqn\IIx
 $$
This gives six distinct fields with dimensions $\{0,{1\over{10}},{3\over5},
{3\over2},{7\over{16}},{3\over{80}}\}$. The simplest primary fields in a
\tci\ model are the products of primary fields in each TCI factor.  The
following table summarizes their dimensions:
 $$\vbox{\offinterlineskip
  \halign{\hfil\ #\ \hfil & \hfil\ #\quad\hfil & \vrule# &
  \hfil\quad#\ \hfil &
  \hfil\ #\ \hfil & \hfil\ #\ \hfil & \hfil\ #\ \hfil &
  \hfil\ #\ \hfil & \hfil\ #\ \hfil \cr
  {} & $T_1$ && $\ph111$ & $\ph112$ &
  $\ph113$ & $\ph114$ & $\ph121$ & $\ph122$ \cr
  \omit&\omit&height4pt&\omit&\omit&\omit&\omit&\omit&\omit\cr
  $T_2$ & + && $0$ & ${1\over{10}}$ & ${3\over5}$ &
  ${3\over2}$ & ${7\over{16}}$ & ${3\over{80}}$ \cr
  \omit&\omit&height4pt&\omit&\omit&\omit&\omit&\omit&\omit\cr
  \noalign{\hrule}
  \omit&\omit&height4pt&\omit&\omit&\omit&\omit&\omit&\omit\cr
  $\ph211$ & $0$ && $0$ & ${1\over{10}}$ &
  ${3\over5}$ & ${3\over2}$ & - & - \cr
  \omit&\omit&height4pt&\omit&\omit&\omit&\omit&\omit&\omit\cr
  $\ph212$ & ${1\over{10}}$ && ${1\over{10}}$ &
  ${1\over5}$ & ${7\over{10}}$ & ${8\over5}$ & - & - \cr
  \omit&\omit&height4pt&\omit&\omit&\omit&\omit&\omit&\omit\cr
  $\ph213$ & ${3\over5}$ && ${3\over5}$ & ${7\over{10}}$ &
  ${6\over5}$ & ${{21}\over{10}}$ & - & - \cr
  \omit&\omit&height4pt&\omit&\omit&\omit&\omit&\omit&\omit\cr
  $\ph214$ & ${3\over2}$ && ${3\over2}$ &
  ${8\over5}$ & ${{21}\over{10}}$ & $3$ & - & - \cr
  \omit&\omit&height4pt&\omit&\omit&\omit&\omit&\omit&\omit\cr
  $\ph221$ & ${7\over{16}}$ && - & - & - & - &
  ${7\over8}$ & ${{19}\over{40}}$ \cr
  \omit&\omit&height4pt&\omit&\omit&\omit&\omit&\omit&\omit\cr
  $\ph222$ & ${3\over{80}}$ && - & - & - & - &
  ${{19}\over{40}}$ & ${3\over{40}}$ \cr}}
  \eqn\IIxi
 $$
The dashes in the table represent dimensions of fields that do not appear
in the \zk8. Their decoupling must be explicitly demonstrated, which we
will do later on, but for now we will simply ignore them.  There exist
infinitely many other primary fields formed {}from appropriate combinations
of descendant fields {}from each TCI factor.  For example, $\ph112\partial
\ph212-[\partial\ph112]\ph212$ is a primary field of dimension $6\over5$.
In general, these more complicated primary fields will have dimensions
differing {}from those given in \IIxi\ by the addition of a positive integer.

Now we turn to the dimensions of the fields in the \zk8 model. The
dimensions $h^j_m$ of the basic set of Virasoro primary fields are given by
eq.~\IIii.  As noted earlier, the full set of primaries of the PF theory
have dimensions of the form $h^j_m +\,${\it positive integers.}  Examining
\IIii\ one finds that the half-integral spin fields, $j\in{\bf Z} +
{1\over2}$, have no counterpart in the \tci\ model, so that we will only be
able to find a representation for the integral spin fields.  As these form a
closed algebra among themselves, the absence of the half-integral spins is
self-consistent.  In fact we will not need these half-integral spin
fields to construct the $K=8$ supercurrents.  Furthermore, only integral spin
fields enter in the $K=8$ fractional superstring partition function [\ArT].
The dimensions of the integral spin fields are summarized in the following
table.
 $$\vbox{\offinterlineskip
  \halign{\hfil#\hfil\ &\vrule#&
  \ \ \hfil\ \ #\ \ \hfil&\hfil\ \ #\ \ \hfil&
  \hfil\ \ #\ \ \hfil&\hfil\ \ #\ \ \hfil&\hfil\ \ #\ \ \hfil\cr
  4&&2&${{15}\over8}$&${3\over2}$&${7\over8}$&0\cr
  \omit&height5pt&\omit&\omit&\omit&\omit&\omit\cr
  3&&${6\over5}$&${{43}\over{40}}$&${7\over{10}}$&${3\over{40}}$&\cr
  \omit&height5pt&\omit&\omit&\omit&\omit&\omit\cr
  2&&${3\over5}$&${{19}\over{40}}$&${1\over{10}}$&&\cr
  \omit&height5pt&\omit&\omit&\omit&\omit&\omit\cr
  1&&${1\over5}$&${3\over{40}}$&&&\cr
  \omit&height7pt&\omit&\omit&\omit&\omit&\omit\cr
  0&&0&&&&\cr
  \omit&height5pt&\omit&\omit&\omit&\omit&\omit\cr
  \noalign{\hrule}
  \omit&height3pt&\omit&\omit&\omit&\omit&\omit\cr
  $\matrix{j&\cr&m\cr}$&&0&1&2&3&4\cr}}
  \eqn\IIxii
 $$
Notice the pleasing fact that every field in the \tci\ table has at least
one partner in the \zk8 table (mod 1) and vice-versa.

Next we will check the equivalence of the characters of the \tci\ model
and the \zk8 theory.  The characters $\chi_{r,s}(q)$ corresponding to the
primary fields $\Phi_{r,s}$ of the TCI model are [\Rocha]
 $$\eqalign{
  \eta(q)\chi_{r,s}(q)&=q^{1\over{24}}\sum_{n\in{\bf Z}}
  (q^{\as n^-}-q^{\as n^+})~,\cr
  \as n^\pm&={{(40n+5r\pm4s)^2-1}\over80},\cr}
  \eqn\IIxiii
 $$
where $\eta(q)$ is the Dedekind $\eta$-function, and $q={\rm e}^{2\pi
i\tau}$ where $\tau$ is the modular parameter of the torus.  The
characters ${\cal Z}^j_m(q)$ for the \zk8 sectors $[\phi^j_m]$ are related
to the known string functions $c^{2j}_{2m}$ by ${\cal Z}^j_m(q) = \eta(q)
c^{2j}_{2m}(q)$ [\Kac].  This gives the expression for the \zk8 characters
 $$\eqalign{
  \eta^2(q){\cal Z}^j_m(q)=&~q^{h^j_m+{1\over40}}
  \sum_{u,v=0}^\infty (-1)^{u+v}
  q^{{1\over2}u(u+1)+{1\over2}v(v+1)+9uv}\cr
  &~~~\times\left[q^{u(j+m)+v(j-m)}-
  q^{9-2j+u(9-j-m)+v(9-j+m)}\right]~.\cr}
  \eqn\IIxiv
 $$
The PF characters satisfy the identities ${\cal Z}^j_m
= {\cal Z}^j_{m+8} = {\cal Z}^{4-j}_{m-4}$ by virtue of the PF field
identifications \IIiii.  Comparing these character formulas, we find the
following nine identities between the \zk8 and \tci\ characters:
 $$\eqalign{
  {\cal Z}^0_0+{\cal Z}^4_0 & = [\chi^2_0+\chi^2_{3\over2}]
   q^{-{c\over24}} \cr
  {\cal Z}^1_0+{\cal Z}^3_0 & = [\chi^2_{1\over10}+\chi^2_{3\over5}]
   q^{-{c\over24}} \cr
  {\cal Z}^2_0 & = [\chi_0\chi_{3\over5}+\chi_{1\over10}\chi_{3\over2}]
   q^{-{c\over24}} \cr
  {\cal Z}^0_1+{\cal Z}^4_1 & = [\chi^2_{7\over16}]
   q^{-{c\over24}} \cr
  {\cal Z}^1_1+{\cal Z}^3_1 & = [\chi^2_{3\over80}]
   q^{-{c\over24}} \cr
  {\cal Z}^2_1 & = [\chi_{7\over16}\chi_{3\over80}]
   q^{-{c\over24}} \cr
  {\cal Z}^0_2 & = [\chi_0\chi_{3\over2}]
   q^{-{c\over24}} \cr
  {\cal Z}^1_2 & = [\chi_{1\over10}\chi_{3\over5}]
   q^{-{c\over24}} \cr
  {\cal Z}^2_2 & = [\chi_0\chi_{1\over10}+\chi_{3\over5}\chi_{3\over2}]
   q^{-{c\over24}} \cr }
  \eqn\IIxv
 $$
for central charge $c={7\over5}$.  This makes the identification of the
\zk8 fields with the \tci\ fields more explicit.

The final check on the equivalence of these two theories
would be to show that the \zk8 fusion rules \IIiv\ are the same as
those for the \tci\ model.  That this is indeed the case will be made
clear in the subsequent discussion [see eqs.~(4.3)-(4.5)].

Notice that in the above identification of \zk8 fields as \tci\ fields, the
PF characters only appear in the particular combinations ${\cal Z}^j_m +
{\cal Z}^{4-j}_m$. These are precisely the combinations that appear in the
modular invariant partition function of the $K=8$ fractional superstring
[\ArT].

\chapter{Solving the TCI associativity constraints}

We present the associativity constraints, following ref.~[\DoFa],
of the chiral TCI model four-point functions and use them
to construct the TCI operator product algebra. In the next section
we expand this discussion to classify the \tci\ operator product
algebras.

What is needed to construct the TCI chiral algebra is a complete
list of the associative transformation
properties of all non-vanishing four-point correlation functions.
That is, for a given four-point function
 $$
  \ba\phi_i(z_i)\phi_j(z_j)\phi_k(z_k)\phi_l(z_l)\fa,
  \eqn\IIIi
 $$
we need to know the transformation matrix between the conformal blocks
as $z_i\rightarrow z_j~(z_k\rightarrow z_l)$ and as $z_j\rightarrow
z_k~(z_l\rightarrow z_i)$.  These matrices are called fusion matrices and
are denoted $\alpha$.  In general, determining $\alpha$ for \IIIi\ is
a difficult and as yet unsolved problem, but for the minimal models in
general and the TCI model in particular it can be solved using the
Feigin-Fuchs technique.  Dotsenko and Fateev [\DoFa] show how to construct
the fusion matrices in these cases.  In Appendix A we review and summarize
the Feigin-Fuchs technique and use it to construct the $\alpha$ matrices
for the TCI model.

The fusion rules of the TCI model are [\BPZ]
 $$\eqalign{
  &\matrix{
  \fpi12\times\fpi12\sim\fpi11+\fpi13 \hfill
  & \fpi12\times\fpi21\sim\fpi22 \hfill \cr
  \fpi12\times\fpi13\sim\fpi12+\fpi14 \hfill
  & \fpi13\times\fpi21\sim\fpi22 \hfill \cr
  \fpi12\times\fpi14\sim\fpi13 \hfill
  & \fpi14\times\fpi21\sim\fpi21 \hfill \cr
  \fpi13\times\fpi13\sim\fpi11+\fpi13\qquad
  & \fpi12\times\fpi22\sim\fpi21+\fpi22\qquad \cr
  \fpi13\times\fpi14\sim\fpi12 \hfill
  & \fpi13\times\fpi22\sim\fpi21+\fpi22\hfill \cr
  \fpi14\times\fpi14\sim\fpi11 \hfill
  & \fpi14\times\fpi22\sim\fpi22 \hfill \cr}\cr
  &\cr
  &\hfill\matrix{
  \fpi21\times\fpi21\sim \fpi11+\fpi14\hfill\cr
  \fpi21\times\fpi22\sim \fpi12+\fpi13\hfill\cr
  \fpi22\times\fpi22\sim\fpi11+\fpi12+\fpi13+\fpi14\hfill\cr}\hfil\cr}
 \eqn\IIIii
 $$
These fusion rules are maximal in the sense that no
additional fields may appear on the right-hand side of a given OPE; on the
other hand, some fields may decouple or have vanishing
structure constants and thereby {\it not}
appear.  The task of finding an associative solution to the chiral TCI
model is a straightforward if laborious one: apply the associativity
constraints of Appendix A to all non-vanishing four-point functions to
determine a consistent set of structure constants to insert into the
fusion rules above.  Additionally, the multiplicity of the fields
must be left free a priori and determined by the fusion rules and
associativity constraints. For example, in the \zk{K\ge3} theory
there are two distinct dimension $1-{1\over K}$ operators, namely
$\psi_1$ and $\psi_{-1}$.

We consider only diagonalizable algebras so that we are free to choose
the normalization
 $$
  q_{ij} = \delta_{ij}
  \eqn\IIIiii
 $$
in \Iiii. In other words, the OPE of any field with itself closes on the
identity with coefficient unity. This normalization implies some symmetries
among the structure constants which we will use. First of all, associativity
of three-point functions implies that structure constants are cyclically
symmetric. Specifically, the following two ways of expanding the same
three-point function must give the same result,
 $$\matrix{
  \ba\phi_i\phi_j\phi_k\fa = c_{ijk}\ba[\phi_k]\phi_k\fa = c_{ijk}, \cr
  \ba\phi_i\phi_j\phi_k\fa = c_{jki}\ba\phi_i[\phi_i]\fa = c_{jki}. \cr}
  \eqn\IIIiv
 $$
Second of all, because the models we are working with are unitary the
interchange of two indices effects the action of complex conjugation so
that we have
 $$
  c_{ijk}=c_{jki}=c_{kij}=c^*_{jik}=c^*_{ikj}=c^*_{kji}.
  \eqn\IIIv
 $$
Using these symmetries we can proceed to calculate the structure constants
for the chiral TCI algebra.  (Note that the only normalization freedom
left after imposing \IIIiii\ is in the sign of the fields.  Thus the
structure constants derived below will only be fixed up to possible signs.)

We will not explicitly show the entire construction, but as an illustrative
example we will show that the $\fpi21$ field decouples {}from the chiral TCI
algebra, and then we will simply state the final result. Therefore, consider
the OPE of the $\fpi21$ field with itself
 $$
  \fpi21\fpi21=\fpi11+c_{(2,1)(2,1)(1,4)}\fpi14~,
  \eqn\IIIvi
 $$
and the four-point function
 $$
  \ba\fpi14\fpi21\fpi14\fpi21\fa.
  \eqn\IIIvii
 $$
This four-point function has only one conformal block and its (one by one)
fusion matrix is $\alpha = -1$ (see Appendix A).  Thus the associativity
constraint derived {}from \IIIvii\ is
 $$
  -c^2_{(1,4)(2,1)(2,1)}=c^2_{(2,1)(1,4)(2,1)}~.
  \eqn\IIIviii
 $$
As the structure constants are cyclically symmetric this implies that
$c_{(2,1)(2,1)(1,4)}=0$. Checking to see if this result is consistent we
look at the four-point function consisting of all $\fpi21$ fields, \ie\
 $$
  \ba\fpi21\fpi21\fpi21\fpi21\fa.
  \eqn\IIIix
 $$
The $\alpha$ matrix for this four-point function [{}from table (A8)]
gives the following associativity constraint
 $$
  {1\over{\sqrt2}}\mtt{1}{8\over7}{7\over8}{-1}
  \pmatrix{1\cr c^2_{(2,1)(2,1)(1,4)}\cr} =
  \pmatrix{1\cr c^2_{(2,1)(2,1)(1,4)}\cr}.
  \eqn\IIIx
 $$
However, plugging the value $c_{(2,1)(2,1)(1,4)}=0$ into this matrix
equation gives a contradiction which is resolved only by decoupling the
$\fpi21$ field {}from the chiral TCI algebra. Similar reasoning shows that
the $\fpi22$ field must also decouple.

In the preceding argument we implicitly took the multiplicity of the
$\fpi21$ field to be one. The argument, however, can be repeated assuming
$n$ copies of $\fpi21$ and it is not difficult to see that the result is the
same: all $n$ $\fpi21$ fields must decouple.

A solution does exist though for the remaining four fields, and a little
work yields the operator product algebra (here $z$ and $w$ dependences
as well as Virasoro descendants are suppressed)
 $$
  \eqalign{
  \fpi12\fpi12 & = \fpi11 - s\fpi13, \cr
  \fpi12\fpi13 & = - s\fpi12 + u\fpi14, \cr
  \fpi12\fpi14 & = u\fpi13, \cr
  \fpi13\fpi13 & = \fpi11 + s\fpi13, \cr
  \fpi13\fpi14 & = u\fpi12, \cr
  \fpi14\fpi14 & = \fpi11, \cr }
  \eqn\IIIxi
 $$
where the structure constants $u$ and $s$ are given by
 $$
  u = \sqrt{3\over7}~,\qquad s = \sqrt{{2\over3}rx}~,
  \eqn\IIIxii
 $$
where
 $$
  r={{\sqrt5-1}\over2}~, \qquad
  x={{\Gamma^2\left({2\over5}\right)}\over
  {\Gamma\left({1\over5}\right)\Gamma\left({3\over5}\right)}}~.
  \eqn\IIIxiii
 $$
Again, we find that increasing the multiplicity of the $\fpi1i$ fields
does not yield any new associative solutions, so that \IIIxi\ is the
only chiral associative TCI algebra.

The new $K=8$ algebra, \ie~the \hh\ FSCA, can be built essentially
{}from the direct tensoring of two TCI algebras given above
plus a free boson. We construct this \hh\ FSCA in Sects.~5 and 6.
However, solving a \tci\ associative
algebra introduces additional solutions. The technical details and
subtleties of this prodedure, which will be needed at
the end of Sect.~6 to construct screening charges, is presented
in the next section.

\chapter{Solving the \tci\ associativity constraints}

In this section we construct the associative solutions to the
\tci\ operator product algebra.  This involves an extended
technical discussion.  However, for the purpose of constructing
the $K=8$ FSCAs, only the two \tci\ solutions, denoted $\ff1$ and
$\ff2$, are relevant.  We use $\ff1$ to construct the \hh\ algebra
and screening charges and $\ff2$ to do the same for the \tj8 FSCA.
In particular, the results for the $\ff1$ algebra OPEs needed to
construct the new \hh\ FSCA are given below in eqs.~(4.13)-(4.15).

We now consider solving the associativity constraints on the \tci\
operator product algebra. The obvious solution is the tensor product
algebra constructed by simply multiplying the associative TCI algebra
found in the last section with another
copy of itself.  We call this algebra $\pp1$. However, solving the
associativity constraints for a tensor product algebra also yields new
solutions which cannot be expressed as the product of the
structure constants of two simple algebras. This must be the case for
the \tci\ model, because at the level of solving a consistent associative
operator product algebra, the mathematics of constructing a chiral tensor
product of two TCI models ($c={7\over10}+{7\over10}={7\over5}$,
${\bar c =0}$) is the same as that of constructing a non-chiral TCI model
($c={7\over10}$, ${\bar c} = {7\over10}$).  In the latter case, we know
there exists a left-right symmetric model (see for example [\Qiu]) in
which the magnetic spin operator, with $(h,{\bar h}) = \left({3\over80},
{3\over80}\right) \sim \fpi22(z)\fpi22 ({\bar z})$, enters.  Such a field
does not appear in the simple product of two chiral TCI models.

We are, of course, interested in the former problem---constructing a
chiral tensor product of two TCI models (\tci)---because our objective
is to find a representation of the $c={7\over5}$, \zk8 theory.  To this end
consider two TCI models with energy-momentum tensors $T_1$ and
$T_2$, (where $T_{Z8}=T_1+T_2$), whose primary fields are $\ph1ij$
and $\ph2ij$ respectively. It is convenient to define the following fields
in the tensor product model
 $$\eqalign{
  \B{ij} & \sim \ph11i\ph21j~, \cr
  \la{ij}~, \lb{ij} & \sim \ph12i\ph22j~. \cr}
  \eqn\IVi
 $$
These identifications are only symbolic since the structure constants
for the $\B{ij}$, $\la{ij}$ and $\lb{ij}$ fields are generically {\it not}
the product of structure constants for the $\fpi ij$ fields, as we have
already discussed.  Note especially that the definition for the $\la{ij}$
fields is identical to that for the $\lb{ij}$ fields. This is because,
as could be guessed {}from the presence of order and
disorder fields with the same conformal dimensions
in the non-chiral, left-right diagonal, TCI model, the $\lambda$ fields
naturally split. In other words, the $\lambda$ fields have multiplicity 2.
There are in fact solutions where the $\lambda$ fields have multiplicity 1,
but these turn out to be simple subalgebras of the larger operator algebras
that we will construct with both $\lambda$ and $\bar\lambda$ fields.
Further increasing the multiplicities of the $\beta$ and $\lambda$,
$\bar\lambda$ fields does not yield any new solutions.

A priori, the ``crossed'' fields $\ph11i\ph22j$ and $\ph12i\ph21j$
must also be considered, but arguments similar to those that ruled out the
existence of the $\fpi21$ and $\fpi22$ fields in a chiral TCI model are
applicable here also. The natural, and mathematically forced, exclusion
of the ``crossed'' fields is actually reassuring since they have no
counterpart in the \zk8 theory that we are trying to represent.

The \tci\ operator algebra for the 24 simple primary fields $\B{ij}$,
$\la{ij}$ and $\lb{ij}$ contains several hundred distinct structure
constants allowed by the fusion rules, although many of them will
turn out to be zero.  As mentioned before, the \tci\ model contains
an infinite number of additional fields which are composed of
descendants of the
simple primaries with respect to $T_1$ or $T_2$, but are primary
with respect to the full energy-momentum tensor $T_1+T_2$.  Thus,
using the conformal Ward identities for each TCI factor separately
($T_1$ or $T_2$), we can derive the structure constants of any \tci\
primary {}from those of the simple primaries \IVi.  This is a crucial
simplifying feature, for it allows us to solve the associativity
constraints for the \tci\ theory by looking only at a finite operator
algebra.

A useful way of organizing the 24 simple primary
fields is summarized in the following table.
 $$\vbox{\offinterlineskip
  \halign{#& \quad#\quad & \vrule# & \quad # & # & # & # & # & # \cr
  $1$&$F_4$&&$\B{33}$&$\B{22}$&$\B{32}$&$\B{23}$
    &$\la{22}$&$\lb{22}$\cr
  $2$&$F_3$&&$\B{31}$&$\B{24}$&$\B{34}$&$\B{21}$
    &$\la{21}$&$\lb{21}$\cr
  $2$&$F_2$&&$\B{13}$&$\B{42}$&$\B{12}$&$\B{43}$
    &$\la{12}$&$\lb{12}$\cr
  $0$&$F_1$&&$\B{11}$&$\B{44}$&$\B{14}$&$\B{41}$
    &$\la{11}$&$\lb{11}$\cr
  \omit&\omit&height2pt&\omit&\omit&\omit&\omit&\omit&\omit\cr
  \noalign{\hrule}
  \omit&\omit&height2pt&\omit&\omit&\omit&\omit&\omit&\omit\cr
  $j$&{}&&$\beta^{(1)}$&$\beta^{(2)}$&$\beta^{(3)}$&$\beta^{(4)}$
    &$\lambda$&${\bar\lambda}$\cr
  \omit&\omit&height9pt&\omit&\omit&\omit&\omit&\omit&\omit\cr
  {}&$|m|$&&$0$&$0$&$2$&$2$&$1$&$1$\cr}}
  \eqn\IVii
 $$
In this table $F_i$ refers to all the fields in the corresponding row and
$\beta^{(i)}$, $\lambda$ and ${\bar \lambda}$ to all fields the
corresponding columns.  The potential usefulness of this grouping is
made apparent by the fact that the fields in these rows and columns
correspond, by the identifications made in Sect.~2, to \zk8
fields with $j$ and $m$ quantum numbers given above.  Note that we
have used the PF field identifications \IIiii\ to restrict $j$ and $|m|$
(though not necessarily simultaneously) to the set $\{0,1,2\}$.
Indeed, {}from the fusion rules \IIIii\
we can summarize the fusion rules for the \tci\ fields neatly in
terms of fusion rules for the row and column labels. Specifically,
we have the $F_i\times F_j$ fusion rules
 $$\matrix{
  F_1\times F_i\sim F_i\hfill
  &F_3\times F_3\sim F_1+F_3\hfill\cr
  F_2\times F_2\sim F_1+F_2\quad
  &F_3\times F_4\sim F_2+F_4\hfill\cr
  F_2\times F_3\sim F_4\hfill
  &F_4\times F_4\sim F_1+F_2+F_3+F_4\cr
  F_2\times F_4\sim F_3+F_4\quad
  &{}\cr}
  \eqn\IViii
 $$
Similarly, the $\beta^{(i)}\times\beta^{(j)}$ fusion rules
can be summarized by all permutations of the five fusion rules
 $$\eqalign{
  \beta^{(1)}\times\beta^{(i)}&\sim\beta^{(i)},\cr
  \beta^{(2)}\times\beta^{(3)}&\sim\beta^{(4)}.\cr}
  \eqn\IViv
 $$
Now in general, {}from \IIIii\ we have $\lambda\times\lambda,\
\lambda\times\bar\lambda,\ \bar\lambda\times\bar\lambda
\sim \beta^{(1)}+\beta^{(2)}+\beta^{(3)}+\beta^{(4)}$. It turns out,
however, that in all general solutions we can define new $\lambda$
and $\bar\lambda$ fields (by taking linear combinations of the old
fields) so that they have the fusion rules
 $$\eqalign{
  \lambda\times\lambda\sim&~\bar\lambda\times\bar\lambda
   \sim\beta^{(1)}+\beta^{(2)},\cr
  &~\lambda\times\bar\lambda\sim\beta^{(3)}+\beta^{(4)}.\cr}
  \eqn\IVv
 $$
The fusion rules of the $\beta^{(i)}$ with $\lambda$ or $\bar\lambda$
follow {}from \IVv.  To calculate the fusion of any two specific simple
primary fields in \IVii, we can use the intersection of the fusion rules
\IViii\ with \IViv\ and \IVv.

The main result of this section is that the \tci\ algebra has exactly
four distinct (maximal) associative solutions which we call $\pp1$,
$\pp2$, $\ff1$ and $\ff2$.  We will begin by describing these four algebras
and their interrelation.  In particular, we will present a set of simple
rules for calculating the structure constants of any of the algebras
given those of the $\ff1$ algebra.  In eq.~(4.6) below and Appendix B we
list all the $\ff1$ structure constants.  At the end of this section we will
briefly outline the arguments that lead to this classification and
construction of the \tci\ associative operator algebras.

The $\ff1$ operator algebra is the largest one and includes all 24 simple
primary fields in \IVii.  The $\ff2$ algebra contains the 12 fields in the
$F_1$ and $F_4$ rows, \ie\ the $F_2$ and $F_3$ fields
decouple, and is {\it not} a subalgebra of $\ff1$.  Now the $\ff1$ and $\ff2$
algebras each have subalgebras in which only the fields in the
$\beta^{(i)}$ columns appear, \ie\ the $\lambda$ and $\bar\lambda$
fields decouple.  These two subalgebras will be denoted $\widetilde{\pp1}$
and $\widetilde{\pp2}$, respectively.  They are related in a simple
way to the two remaining algebras $\pp1$ and $\pp2$, which, however, are
not subalgebras of any other associative operator product algebras.
In particular, the operator content of $\pp i$ and $\widetilde{\pp i}$ are
the same.  $\pp1$ contains only the 16 $\B{ij}$ fields while $\pp 2$
contains
only the 8 $\B{ij}$ fields in the rows $F_1$ and $F_4$.  As we mentioned
before, the $\pp 1$ algebra is the simple tensor product of two associative
chiral TCI operator algebras, found in the last section.

The interrelationships between these algebras are summarized  in the
diagram
 $$\matrix{
  \pp1&\buildrel {\cal C}\over\longleftrightarrow
    &\widetilde{\pp1}&\subset&\ff1\cr
  \Big\downarrow\rlap{$\vcenter{\hbox{$\scriptstyle{\cal R}$}}$}&{}&
  \Big\downarrow\rlap{$\vcenter{\hbox{$\scriptstyle{\cal R}$}}$}&{}&
  \Big\downarrow\rlap{$\vcenter{\hbox{$\scriptstyle{\cal R}$}}$}\cr
  \pp2&\buildrel {\cal C}\over\longleftrightarrow
    &\widetilde{\pp2}&\subset&\ff2\cr}
  \eqn\IVvi
 $$
The row and column maps, ${\cal R}$ and ${\cal C}$ respectively,
are given by simple rules, and allow us to construct all the
operator algebras starting {}from $\ff1$.  In particular, the column map
${\cal C}$ is given by the rule that we multiply the structure
constants coupling fields in the $\beta^{(2)}$, $\beta^{(3)}$ and
$\beta^{(4)}$ columns by a factor of $-i$, while leaving all other
structure constants the same:
 $$
  {\cal C}:\quad\ba\beta^{(2)}\beta^{(3)}\beta^{(4)}\fa
  \longleftrightarrow
  -i\ba\beta^{(2)}\beta^{(3)}\beta^{(4)}\fa.
  \eqn\IVvii
 $$
The map for these $\beta^{(i)}$ fields in other orderings is determined
by the structure constant symmetries \IIIv.  Note that the sign of $i$
in \IVvii\ is actually of no consequence, since we have not fixed the
signs of the normalizations of the fields; however, the phase $i$ itself
cannot be defined away by any change in the normalization of the fields.

To describe the row map
${\cal R}$, we must first mention that we will always write
the structure constants of the \tci\ operator algebras as functions
of the formal ``variables'' $r$ and $x$.  The numerical value of any such
structure constant is found, however, by letting these ``variables'' take
the values
 $$
  r={{\sqrt5-1}\over2}~, \qquad
  x={{\Gamma^2\left({2\over5}\right)}\over
  {\Gamma\left({1\over5}\right)\Gamma\left({3\over5}\right)}}~,
  \eqn\IVviii
 $$
that weare derived {}from the chiral TCI associativity conditions in Sect.~3.
Consider the $r$ and $x$ dependence of the OPEs of fields in the rows
$F_i$.  For the $\ff1$ and $\pp 1$ algebras, these OPEs are (symbolically)
 $$
  F_i\cdot F_j\sim\sum_k (rx)^{a_k}F_k~,
  \eqn\IVix
 $$
where the exponents $a_k$ can take values
in the set $a_k\in\{0,{1\over2},1\}$.  In other words, the $r$ and $x$
dependence of these structure constants is always proportional to $rx$
raised to one of the above-mentioned powers; furthermore, these powers
are the same for all fields in a given row $F_k$.  The rule for the
row map ${\cal R}$ is now easy to state:  first decouple the $F_2$
and $F_3$ fields and then, treating the different powers $(rx)^{a_k}$
as independent parameters, make the  substitutions
 $$
  {\cal R}:\quad\left\{\matrix{
  (rx)^0\hfill&\rightarrow\ 1~,\hfill\cr
  (rx)^{1/2}\hfill&\rightarrow\ 0~,\hfill\cr
  (rx)^1\hfill&\rightarrow\ {x\over\sqrt r}~.\hfill\cr}\right.
  \eqn\IVx
 $$
For example, in the $\ff1$ algebra the $F_4\cdot F_4$ OPEs \IViii\
can be written symbolically as,
 $$
  F_4\cdot F_4\sim F_1+\sqrt{rx}F_2+\sqrt{rx}F_3+rxF_4,
  \eqn\IVxi
 $$
where the factors $\sqrt{rx}$ and $rx$ indicate the dependence on
$r$ and $x$ in their respective structure constants.  The row map
\IVx\ then leads to the $\ff2$ OPEs
 $$
  F_4\cdot F_4\sim F_1+{x\over\sqrt r}F_4,
  \eqn\IVxii
 $$
where all other numerical factors in the structure constants remain
the same.

We will now write down the structure constants of the $\ff1$ operator
algebra.  To start with, the OPEs of the $\beta^{(i)}$ fields (\ie\ the
$\widetilde{\pp1}$ subalgebra) can be succinctly written as
 $$
  {\cal F}1:\qquad\B{ij}^{(r)}\B{kl}^{(s)}=
  \sum_{t=1}^4 d_{rst}\sum_{m,n=1}^4
  c_{ikm}c_{jln}\B{mn}^{(t)},
  \eqn\IVxiii
 $$
where $c_{ijk}$ and $d_{rst}$ obey the structure constant symmetries
\IIIv. Their non-zero components have the values,
 $$\eqalign{
  c_{1ii}=&~1,\cr
  c_{333}=-c_{223}=&~\sqrt{{2\over3}rx}~,\cr
  c_{234}=&~\sqrt{3\over7},\cr}
  \eqn\IVxiv
 $$
and
 $$\eqalign{
  d_{1rr}&=1~,\cr
  d_{234}&=i~.\cr}
  \eqn\IVxv
 $$
Note that the $c_{ijk}$ are just the chiral TCI structure constants
derived in Sect.~3.  Indeed, if we operate with the column map
\IVvii\ on these structure constants, we obtain the $\pp 1$ operator
algebra structure constants.  This has the effect of letting
$d_{rst}\rightarrow d_{rst}'$ in \IVxiii, where
 $$
  d_{1rr}'=d_{234}'=1~.
  \eqn\IVxvi
 $$
The inclusion of the column label superscripts and $\sum_{t=1}^4 d_{rst}'$
in \IVxiii\ becomes redundant in this case, so that we obtain
 $$
  \pp1:\qquad\B{ij}\B{kl}=\sum_{m,n=1}^4 c_{ikm}c_{jln}\B{mn}~,
  \eqn\IVxvii
 $$
which is just the direct product of two chiral TCI associative solutions,
as advertised.

It remains to include the fields in the $\lambda$ and $\bar\lambda$
columns in the $\ff1$ algebra.  The couplings of these fields are found by
solving the associativity constraints, as discussed below.  A unique
maximal solution is found; a compilation of the resulting structure
constants is given in Appendix B.  The other maximal \tci\ operator
algebras can be found by acting with the ${\cal R}$ and ${\cal C}$
maps as outlined above \IVvi.  The $\ff2$ operator algebra found by
this procedure has structure constants that are equivalent (up to
normalizations) to those found by Qiu in ref.~[\Qiu] for the non-chiral
left-right symmetric TCI model.

So far we have simply described the four inequivalent associative
\tci\ operator algebras.  We will now briefly describe what is
involved in solving the \tci\ associativity constraints.  The basic
observation is that, since the simple primary fields of the \tci\
model are just products of the primaries of each TCI factor, the
fusion matrices for the four-point correlators of the \tci\ simple
primaries are simply the tensor product of the two fusion matrices
for the associated four-point function for each TCI factor separately.
Let us give an example which illustrates this point.  Consider the
$\ba\B{33}\B{33}\B{33}\B{33}\fa$ four-point function.  Now, {}from
\IVi\ $\B{33}=\ph113\ph213$, where the superscripts
denote the two TCI factors.  The general $\fpi 13\fpi 13$ OPE in
the chiral TCI model is, \IIIii,
 $$
 \fpi13\fpi13=\fpi11 + c_{333}\fpi13 ~.
 \eqn\IVxviii
 $$
The associativity condition {}from the $\ba\fpi13\fpi13
\fpi13\fpi13\fa$ four-point function is [{}from eq.~(A8) in
Appendix A] the eigenvalue equation
 $$
  \pmatrix{r&{3\over2}{r\over x}\cr{2\over3}x&-r\cr}
  \pmatrix{1\cr c^2_{333}\cr}=\pmatrix{1\cr c^2_{333}\cr},
  \eqn\IVxix
 $$
which has the unique solution
 $$
  c^2_{333}={2\over3}rx~.
  \eqn\IVxx
 $$
The fact that the matrix in \IVxix\ has exactly one $+1$ eigenvalue is
at the heart of the uniqueness of the single chiral TCI model. However,
when this matrix is tensored with itself to find the associativity
constraint for the $\ba\B{33}\B{33}\B{33}\B{33}\fa$ four-point
function, we find that it now has two $+1$ eigenvalues which implies
that it has a line of solutions.  One of these solutions, of course, is just
the direct product of the chiral TCI solution, giving the $\B{33}\B{33}$
OPE
 $$
  \B{33}\B{33}=\B{11}+c_{333}\B{13}+
  c_{333}\B{31}+c_{333}^2\B{33}~,
  \eqn\IVxxi
 $$
for $c_{333}$ given by \IVxx.  When the associativity constraints of
additional four-point functions are considered\foot{A set of four-point
functions sufficient to show that \IVxxi\ is the only consistent solution
is, for example: $\ba \B{13} \B{33} \B{33} \B{33} \fa$, $\ba \B{13} \B{33}
\B{13} \B{33} \fa$ and $\ba \B{13} \B{33} \B{31} \B{33} \fa$.},
{\it without\/} decoupling the $\B{13}$ and $\B{31}$ fields (\ie, for
solutions in which these two fields have multiplicity one, not zero), we
find that only the solution \IVxxi, out of
the line of possible solutions, is
consistent.  (This solution corresponds to either the $\pp 1$
or $\ff1$ operator
algebras, since the $\B{33}\B{33}$ OPE happens to be the same for
both algebras.)  Decoupling $\B{13}$ and $\B{31}$ leads to another solution
on the line---the only other consistent solution---which is a
result of the eigenvalue equation
 $$
  \pmatrix{r^2&{3r^2\over2x}&{3r^2\over2x}&{9r^2\over4x^2}\cr
    {2\over3}xr&-r^2&r&-{3r^2\over2x}\cr
    {2\over3}xr&r&-r^2&-{3r^2\over2x}\cr
    {4\over9}x^2&-{2\over3}xr&-{2\over3}xr&r^2\cr}
  \pmatrix{1\cr0\cr0\cr {4x^2\over9r}\cr}=
  \pmatrix{1\cr0\cr0\cr {4x^2\over9r}\cr}.
  \eqn\IVxxii
 $$
This leads to the OPE
 $$
  \B{33}\B{33}=\B{11}+{2\over3}{x\over\sqrt r}\B{33}.
  \eqn\IVxxiii
 $$
This OPE is part of the $\ff2$ and $\pp2$ operator algebras;  it is easy to
check that \IVxxi\ and \IVxxiii\ are related by the ${\cal R}$ map.

This example illustrates how, even though the chiral TCI model
has a single solution to its associativity constraints, the tensor
product of two such models may have multiple solutions.  Indeed,
the pattern illustrated in the above example for the $\beta^{(1)}$ fields
extends also to the fields in the other columns $\beta^{(i)}$, $\lambda$
and $\bar\lambda$.  To show that the complete set of solutions to the
associativity conditions for the \tci\ operator product algebra is
$\ff1$, $\ff2$, $\pp 1$, $\pp 2$ and their subalgebras,
requires the systematic
and unilluminating examination of all the four-point associativity
constraints.

This completes the construction of the four maximal \tci\ operator
algebras.  We use $\ff1$ (or $\pp 1$) to construct
the \hh\ FSCA and its associated screening charges in
Sects.~5 and 6. In Sect.~7 we
show how the $\ff2$ (or $\pp 2$) algebra leads to the \tj8 FSCA that was
constructed in ref.~[\agt].

\chapter{Construction of the \hh\ FSCA for $c={12\over5}$}

In this section we will construct the set of $K=8$ FSCA currents
\hh\ and derive their structure constants for the special value of
the central charge $c={{12}\over5}$.  As explained in the introduction,
at this value  of the central charge the $K=8$ FSCA is embedded in
the tensor product of the \tci\ model with a free boson.

We set the normalization of the boson $\varphi(z)$ by $\ba\varphi(z)
\varphi(w)\fa=+{\rm ln}(z-w)$ so that its energy-momentum tensor
is
 $$
  T_\varphi(z) = +{1\over2}:\pv(z)\pv(z):~.
  \eqn\Vi
 $$
The dimension 1 primary field $\pv(z)$ will play a special role in
what follows.  Its OPEs are
 $$\eqalign{
  T_\varphi(z)T_\varphi(0) &= {1\over2}z^{-4}
    +2z^{-2}T_\varphi(0)+\cdots\cr
  T_\varphi(z)\pv(0) &= z^{-2}\pv(0)+z^{-1}\pvp2(0)+\cdots\cr
  \pv(z)T_\varphi(0) &= z^{-2}\pv(0)+0+\cdots\cr
  \pv(z)\pv(0)&=z^{-2}+2T_\varphi(0)+\cdots\ .\cr}
  \eqn\Vii
 $$
The full energy-momentum tensor for the \tci\ plus boson theory
is then $T=T_{Z8}+T_\varphi$.  Here the \tci\ (or \zk8) energy-momentum
tensor $T_{Z8}=T_1+T_2$ where the $T_i$ are the energy-momentum
tensors for each TCI factor separately.

To construct our fractional supersymmetry current $J$ we need to
find all dimension ${6\over5}$ fields primary with respect to $T$.
Since we want to generalize the idea of the superconformal current
to one which transforms bosons to {\it parafermion\/} fields, we
impose, as an {\it anzatz}, the condition that only the derivatives
of $\varphi$ enter into the expression for the currents. This means,
in particular, that the vertex operators
${\rm exp}\{\alpha\varphi(z)\}$ need not be
considered.  With this restriction the complete list of dimension
${6\over5}$ primary fields in the \tci\ model tensored with a free
boson consists of the three fields $\B{33}$, $\B{22}\pv$, and
$\B{22}'$.  Here $\B{22}'$ is a \tci\ primary
field made {}from descendant fields {}from each TCI factor:
  $$
  \B{22}'=\partial\ph112\ph212-\ph112\partial\ph212~.
  \eqn\Viia
  $$
Therefore, $J$ is a linear combination of these three fields.

The conditions we impose on this linear combination are that the
$JJ$ OPE close back on itself (\ie\ no other linearly independent
dimension ${6\over5}$ field enters) and that it close on the identity
with coefficient unity.  The calculation of this OPE, however, uses the
structure constants of the \tci\ theory.  Therefore we must first
choose which of the four inequivalent \tci\ operator product algebras,
found in Sect.~4, to use to calculate the $K=8$ FSCA currents.
Now, it so happens that the OPEs of \tci\ fields listed above that
contribute to the $J$ current are left unchanged by the ``column map''
${\cal C}$ introduced in the last section.  This means that the OPEs
for these fields in the $\pp1$ and $\ff1$ \tci\ operator algebras
are the same;  and similarly for the OPEs in the $\pp2$ and $\ff2$
algebras.  Thus, as far as the calculation of the currents is concerned,
there are really only two independent choices of \tci\ algebra to
make.  However, we will later find that only for the $\ff i$ algebras
can we construct appropriate screening charges for the FSCAs when
the boson field $\varphi$ has background charge.  Thus, one should
view the construction of the FSCAs based on the $\pp i$ algebras as
reflecting the existence of ``special'' unitary representations of the
FSCAs for central charge $c={12\over5}$.

In this and the next section, we will construct the $K=8$ FSCAs based
on the $\ff1$ (or $\pp1$) \tci\ operator algebras.  We will find
that the \hh\ algebra, described in the introduction, emerges in this
case.  In Sect.~7 we will construct the \tj8 FSCA which is based on
$\ff2$ (or $\pp2$).

We can now proceed with the determination of the $J$ current.  Using
the $\ff1$ (or $\pp1$) OPEs described in the last section, we can
compute the relevant \tci\ OPEs.  Suppressing all primary fields
but the identity, $\B{22}$, $\B{22}'$ and $\B{33}$ on the
right hand sides of these OPEs, we find
 $$\matrix{
  \B{22}\B{22}=1+s^2\B{33}\hfill
  &\quad\B{22}\B{33}=\B{33}\B{22}=s^2\B{22}\hfill\cr
  \B{22}'\B{22}'
    =-{2\over5}-{4\over5}s^2\B{33}\hfill
  &\quad\B{22}'\B{33}=\B{33}\B{22}'
    =-2s^2\B{22}'\hfill\cr
  \B{33}\B{33}=1+s^2\B{33}\hfill
  &\quad\B{22}\B{22}'=\B{22}'\B{22}
    =0\hfill\cr}
 \eqn\Viib
 $$
where $s=\sqrt{{2\over3}rx}$ was defined in Sect.~3, and where we
have also suppressed the $z$ and $w$-dependence of the fields
and their coefficients in the OPEs.
The only technical difficulty encountered
in computing \Viib\ occurs in the $\B{22}'\B{33}$ OPE.
In this case the TCI Virasoro Ward identities must be used to deduce
the coefficient of the $\fpi12$ descendant that contributes to
$\B{22}'$ \Viia.  In particular, the relevant coefficient
is that of the second term of eq.~(C12) in Appendix C, where the
Virasoro Ward identities are systematically solved for the first
few descendants.  Since $J$ is a linear combination of the
dimension ${6\over5}$ fields mentioned before, and requiring that
the $JJ$ OPE close back on $J$, and on the identity with coefficient
unity, it is easy to see that $J$ is uniquely fixed to be
 $$
  J(z)={1\over\sqrt2}\B{33}(z) + {1\over\sqrt2}\B{22}(z)\pv(z).
  \eqn\Viii
 $$
(Note that $\B{22}'$ does not, in the end, contribute to $J$.)

By keeping track of fields with dimensions other than zero and
$6\over5$ (which correspond to the identity and $J$, respectively)
in the $JJ$ OPE, we can calculate any other currents that couple in
the FSCA.  We find that that the $JJ$ OPE closes on additional
fields with dimensions ${3\over5}$ (mod 1). Since the $\B{33}\B{33}$
and $\B{22}\B{22}$ OPEs both close on the fields $\B{13}$ and
$\B{31}$, with dimensions ${3\over5}$, we might expect that
there are additional currents in the FSCA of dimension ${3\over5}$.
In fact, the ${3\over5}$ fields
all cancel among themselves as do the ${8\over5}$ fields so that the
new FSCA currents have dimensions ${13\over5}$. It turns out that
there are two of them (they could be combined and written as one, but
they naturally split) which we call $\hs1$ and $\hs2$.  Performing the
calculation using the techniques for computing Virasoro descendants
outlined in Appendix C we find the following form for $\hs1$:
 $$\eqalign{
  \hs1(z)=\sqrt{{91}\over{120}}\Biggl[-{{75}\over{142}}
  &\partial^2\B{13}(z)+{{55}\over{71}}
  (L_{-2}\B{13})(z)+{6\over{71}}\B{13}'(z)\cr
  &-\B{13}(z)T_\varphi(z)-\sqrt{3\over7}\B{42}(z)\pv(z)\Biggr]~,\cr}
  \eqn\Viv
 $$
where the field $\B{13}'$, primary with respect to $T_{Z8}$, is defined by
 $$
  \B{13}'(z) = -{{15}\over{26}}\partial^2\ph213(z)
  +{{11}\over{13}}\lph2132(z)-
  {5\over7}T_1(z)\ph213(z).
  \eqn\Vv
 $$
The notation for descendant fields is context dependent.  In particular,
the $L_{-2}$ in \Viv\ refers to the mode of the \tci\ energy-momentum
tensor $T_{Z8}=T_1+T_2$, while the $L_{-2}$ in \Vv\ refers to the mode
of the energy-momentum tensor $T_2$ of the second TCI factor.  Thus,
for example, $(L_{-2}\B{13})=T_1\ph213+(L_{-2}\Phi)^{(2)}_{1,3}$.
The form of $\hs2$ is found by making the substitutions
$\B{13}\rightarrow\B{31}$ and $\B{42}\rightarrow\B{24}$ in \Viv\
and \Vv.  The normalization factor $\sf{91}{120}$ is included to make
the $\hs1\hs1$ OPE close on the identity with coefficient unity.

We can now write down the \hh\ algebra for $c={{12}\over5}$.
Performing some lengthy calculations relying heavily on Appendix
C, we find the operator product algebra
 $$\eqalign{
  JJ&=1+s^2\sqrt2J+s\sf{120}{91}(\hs1+\hs2),\cr
  J\hs1&=s\sf{120}{91}J+{{13}\over{7\sqrt2}}\hs2,\cr
  J\hs2&=s\sf{120}{91}J+{{13}\over{7\sqrt2}}\hs1,\cr
  \hs1\hs1&=1-s\sf{10}{273}\hs1,\cr
  \hs2\hs2&=1-s\sf{10}{273}\hs2,\cr
  \hs1\hs2&={{13}\over{7\sqrt2}}J,\cr}
  \eqn\Vvi
 $$
where recall that $s=\sqrt{{2\over3}rx}$.  In \Vvi\ we have suppressed
the $z$ and $w$ dependences (which can easily be restored) as well as the
Virasoro descendant fields.

There is an important point to make about descendant fields on the
right hand side of the OPEs \Vvi.  In addition to all the Virasoro
descendants, there will be, in general, an infinite number of
Virasoro primaries.
Every new Virasoro primary entering on the right hand side of \Vvi\
will have dimensions 1, ${6\over5}$ or ${13\over5}$ (mod 1).  Since
they add no new cuts to the FSCA OPEs, they can all be considered to be
{\it current algebra\/} descendant fields. We will refer to such a field
whose dimension differs {}from that of the current $J$ by an integer $n$,
as a level-$n$ current algebra descendant of $J$. Identical definitions
apply to the currents 1 and $\hs\ell$.

For instance, consider the $JJ\sim1+\cdots$ OPE. We find explicitly
that
 $$\eqalign{
  &J(z)J(0)=z^{-{12\over5}}\times\cr
  &\qquad\bigg\{\Big[1+z^2T(0)+{z^3\over2}\partial T(0)
  +{z^4\over34}\left\{7L_{-2}T+3\partial^2T\right\}(0)+\cdots\Big]\cr
  &\qquad\qquad+{3z^4\over{119N}}\Big[G(0)+{z\over2}\partial G(0)
  +\cdots\Big]+\cdots\bigg\}+\cdots,\cr}
  \eqn\Vvii
 $$
where $G$ is a dimension 4 Virasoro primary of the following form,
 $$\eqalign{
  G=N\Big[&17\B{44}\pv-{5\over2}(L_{-2}T_1+L_{-2}T_2+2T_1T_\varphi+
  2T_2 T_\varphi)\cr
  &+{{35}\over{54}}L_{-2}T_\varphi-{7\over{36}}\partial^2T_\varphi+
  {{305}\over7}T_1T_2+{3\over4}(\partial^2T_1+\partial^2T_2)\Big].\cr}
  \eqn\Vviii
 $$
Thus $G$ is a level-4 current algebra descendant of the identity.
This field $G$ also enters the $\hs1\hs1$ OPE:
 $$
  \hs1\hs1=1+\cdots+{{39}\over{2380N}}G+\cdots,
  \eqn\Vix
 $$
and similarly for $\hs2\hs2$.  The point is that any given current algebra
descendant, such as $G$, may or may not appear in a given representation
of the \hh\ algebra.  One could decide to include $G$ among the defining
currents of the FSCA, with the effect of reducing the number of
representations of the algebra.  On the other hand, the currents $\hs1$
and $\hs2$ must be included in the definition of the chiral algebra because
they are associated with new cuts in the FSCA OPEs.

\chapter{The \hh\ algebra at arbitrary central charge}

In this section we extend the construction of the \hh\ algebra to
general central charge by adding a background charge to the $\varphi$
boson.  The boson energy-momentum tensor is then
 $$
 T_\varphi(z) = {1\over2}:\pv(z)\pv(z):-~\as0\pvp2(z).
  \eqn\VIi
 $$
With this normalization, the central charge for the boson then becomes,
 $$
  c_\varphi=1-12\as0^2,
  \eqn\VIii
 $$
and the total central charge for the boson plus \tci\ theory is
 $$
  c={{12}\over5}-12\as0^2.
  \eqn\VIiii
 $$
$\pv$ is no longer a primary field so its OPEs with itself and
$T_\varphi$ are more complicated than in \Vii. The OPEs with the
background charge $\as0$ turned on are
 $$\eqalign{
  T_\varphi(z) T_\varphi(0) &= {{1-12\as0^2}\over2}z^{-4}
    +2z^{-2}T_\varphi(0)+\cdots\cr
  T_\varphi(z) \pv(0) &= 2\as0z^{-3}
    +z^{-2}\pv(0)+z^{-1}\pvp2(0)+\cdots\cr
  \pv(z) T_\varphi(0) &= -2\as0z^{-3} + z^{-2}\pv(0)+0+\cdots\cr
  \pv(z)\pv(0)&=z^{-2}+2T_\varphi(0)+2\as0\pvp2(0)+\cdots\ .\cr}
  \eqn\VIiv
 $$
Now we are in a position to construct the $J$, $\hs1$
and $\hs2$ currents for a general central charge $c$.

We start with $J$ and notice right away that as written
in \Viii\ it is no longer primary because $\B{22}\pv$ is
not.  However, using the OPEs \VIiv, it is found that the
following combination is primary
 $$
  \B{22}(z)\pv(z)-5\as0\partial\B{22}(z).
  \eqn\VIv
 $$
Demanding the same conditions that we did when
constructing $J$ for $c={12\over5}$, namely, that the $JJ$
OPE close on the identity with coefficient unity and that it
close on itself without any new dimension ${6\over5}$ fields
entering, the form of $J$ for general $c$ is then fixed to be
 $$\eqalign{
  J(z)=&~{1\over\sqrt2}\sqrt{1\over{1-5\as0^2}}
  \Big[\B{22}(z)\pv(z)-5\as0\partial\B{22}(z)\Big]\cr
  &\qquad+{1\over\sqrt2}\sqrt
  {{1+4\as0^2}\over{1-5\as0^2}}\B{33}(z)~.\cr}
  \eqn\VIvi
 $$

To discover the form for $\hs1$ and $\hs2$
we calculate the OPE of $J$ with itself and pick out
all dimension ${{13}\over5}$ pieces. The fact that the ${3\over5}$
and ${8\over5}$ pieces cancel among themselves and drop out is
certainly not obvious, but explicit calculation demonstrates that this
is indeed the case.  We find
 $$\eqalign{
  \hs1=N_H\Bigg[&-{75\over142}\left(1+{49\over3}\as0^2\right)
  \partial^2\B{13}+{6\over71}\left(1-31\as0^2\right)\B{13}'\cr
  &-\B{13}T_\varphi-{3\over2}\as0\B{13}\pvp2
  +{5\over2}\as0\partial\B{13}\pv\cr
  &+{55\over71}\left(1+{156\over11}\as0^2\right)L_{-2}\B{13}\cr
  &+\sqrt{3(1+4\as0^2)\over7}\left(-\B{42}\pv+{5\over8}\as0
  \partial\B{42}+{7\over8}\as0\B{42}'\right)\Bigg],\cr}
  \eqn\VIvii
 $$
where $\B{42}'$ is the Virasoro primary field
 $$
  \B{42}'(z)=\partial\ph114(z)\ph212(z)-15\ph114(z)\partial\ph212(z).
  \eqn\VIviii
 $$
As before, $\hs2$ is obtained by substituting $\B{13}\rightarrow\B{31}$
and
$\B{42}\rightarrow\B{24}$ in the above formulas. The normalization
constant $N_H$ is determined by calculating the $\hs1\hs1$
OPE and normalizing it to close on the identity
with coefficient unity. Doing this gives,
 $$
  N_H=\sqrt{{91\over120}{1\over(1-5\as0^2)(1-{33\over4}\as0^2)}}.
  \eqn\VIix
 $$

The result for the complete OPEs for $J$,
$\hs1$ and $\hs2$ can now be presented (again suppressing the $z$ and
$w$ dependences)
 $$\eqalign{
  JJ&=1+s\sf{120}{91}
   \sf{1-{{33}\over4}\as0^2}{1-5\as0^2}
   (\hs1+\hs2)+s^2
   \sqrt2\sf{1+4\as0^2}{1-5\as0^2}J,\cr
  &\cr
  J\hs1&=s\sf{120}{91}\sf{1-{{33}\over4}\as0^2}{1-5\as0^2}J
   +{{13}\over{7\sqrt2}}\sf{1+4\as0^2}{1-5\as0^2}\hs2,\cr
  &\cr
  J\hs2&=s\sf{120}{91}\sf{1-{{33}\over4}\as0^2}{1-5\as0^2}J
   +{{13}\over{7\sqrt2}}\sf{1+4\as0^2}{1-5\as0^2}\hs1,\cr
  &\cr
  \hs1\hs1&=1-s\sf{10}{273}
   {{1-{{519}\over5}\as0^2}\over
   {\sqrt{(1-5\as0^2)(1-{{33}\over4}
   \as0^2)}}}\hs1,\cr
  &\cr
  \hs2\hs2&=1-s\sf{10}{273}
   {{1-{{519}\over5}\as0^2}\over
   {\sqrt{(1-5\as0^2)(1-{{33}\over4}
   \as0^2)}}}\hs2,\cr
  &\cr
  \hs1\hs2&={{13}\over{7\sqrt2}}
   \sf{1+4\as0^2}{1-5\as0^2}J.\cr}
  \eqn\VIx
 $$
This is the \hh\ FSCA algebra presented in the introduction, where
\VIiii\ was used to write the structure constants in terms of the
central charge instead of the background charge.

By constructing the FSCA OPEs with arbitrary background charge
we have derived the dependence of the FSCA structure constants
on the central charge.  However, we have not
shown that the \hh\ operator algebra possesses an interesting
representation theory.  In particular, the representation in
which we constructed it, namely \tci$\otimes\{\varphi,\alpha_0\}$
(the second factor representing the boson with
background charge), is generically a nonunitary theory.  To show
that the \hh\ FSCA actually has unitary representations for
$c\neq{12\over5}$, we follow the Feigin-Fuchs approach
[\DoFa,\Felder] of constructing a projection of the
\tci$\otimes\{\varphi,\alpha_0\}$ Fock space to a unitary
subspace.  The crucial step in this procedure is the
construction of ``screening charges,'' dimension zero
operators which commute with all the currents of the algebra.
The screening charges $Q_{\pm}$ can be written as line
integrals of dimension 1 currents $V_{\pm}(z)$
 $$
 Q_\pm = \oint{{\rm d}z\over2\pi i}V_\pm(z)~.
 \eqn\VIxa
 $$
The condition that the screening charges commute with the FSCA
currents is equivalent to the condition that the screening
currents $V_\pm(z)$ have zero (or a total derivative) as the
residue of the single pole term in their OPEs with the currents.

Using the $\ff1$ \tci\ OPEs, we can construct such screening
currents for the \hh\ FSCA.
They are found to be of the form
 $$
  V_+(z)=\la{11}(z)S_+(z),\qquad V_-(z)=\lb{11}(z)S_-(z),
  \eqn\VIxi
 $$
where $S_\pm(z)$ are the dimension ${1\over8}$ fields, primary with
respect to $T_\varphi(z)$ \VIi,
 $$
  S_\pm(z)={\rm e}^{\as\pm\varphi(z)},\qquad
  \as\pm=-\as0\pm\sqrt{\as0^2+{1\over4}},
  \eqn\VIxii
 $$
so that $V_\pm(z)$ are dimension 1 primary fields with respect
to $T(z)$.  Note that these screening currents involve
$\lambda$ and $\bar\lambda$ fields, which do {\it not\/}
appear in the $\pp1$ \tci\ operator algebra, defined in Sect.~4.
Thus, even though the \hh\ algebra at arbitrary background
charge \VIx\ could be constructed in the $\pp1$ model, we find
that its screening charges cannot.

To show that the associated screening charges do indeed
commute with the FSCA currents, we have to evaluate the OPEs
of $V_\pm(z)$ with those currents.  Because $V_\pm(z)$ were
constructed to be Virasoro primary fields of dimension one, they
satisfy the OPEs
 $$
 V_\pm(z)T(w)={V_\pm(w)\over(z-w)^2}+{\rm regular~terms}~,
 \eqn\VIxiia
 $$
and thus $Q_\pm$ automatically commute with $T(z)$.

The $V_\pm$ OPEs with the other currents
are conveniently written in terms of the following fields:
 $$\matrix{
  {1\over5}:\quad&\hfill U_+(z)\ =&\la{22}(z)S_+(z),\hfill\cr
  {3\over5}:\quad&\hfill W^{(1)}_+(z)\ =&\la{12}(z)S_+(z),\hfill\cr
  {8\over5}:\quad&\hfill X^{(1)}_+(z)\ =&\la{12}'(z)S_+(z),\hfill\cr
  {8\over5}:\quad&\hfill Y^{(1)}_+(z)\ =&5\partial\la{12}(z)S_+(z)
   -19\la{12}(z)\partial S_+(z),\hfill\cr}
  \eqn\VIxiii
 $$
where the fraction preceding each charge is its conformal dimension
and we have defined the following field, primary with
respect to $T_{Z8}(z)$,
 $$
  \la{12}'(z)=3\partial\ph121(z)\ph222(z)-35\ph121(z)\partial\ph222(z).
  \eqn\VIxiv
 $$
We define more fields $W^{(2)}_+$, $X^{(2)}_+$ and $Y^{(2)}_+$ by making
the substitution $\la{12}\rightarrow\la{21}$ in \VIxiii; we also define
the fields with minus subscripts letting $S_+\rightarrow S_-$ and
$\lambda\rightarrow\bar\lambda$.
Now, using Appendices B and C, we calculate the relevant OPEs to find
 $$\eqalign{
  &V_\pm(z)J(w)=\pm\sf2{1-5\as0^2}\as\mp{U_\pm(w)\over(z-w)^2}
   +{\rm reg.}\cr
  &\cr
  &V_\pm(z)H_\ell(w)=-i\sf34N_H\bigg[
   -4\as0\as\mp{W_\pm^{(\ell)}(w)\over(z-w)^3}+
   {10\over3}\as0\as\mp{\partial W_\pm^{(\ell)}(w)\over(z-w)^2}\cr
  &~~+{16\over399}(2-15\as0\as\mp){X_\pm^{(\ell)}(w)\over(z-w)^2}
   -{2\over57}(3-32\as0\as\mp){Y_\pm^{(\ell)}(w)\over(z-w)^2}\bigg]
   +{\rm reg.}\cr}
  \eqn\VIxv
 $$
Since the single pole term vanishes in these OPEs, we have confirmed
that the screening charges $Q_\pm$ commute with $J$ and the $H_\ell$.
In fact, {}from Appendix C, it is easy to check that the level 2 Virasoro
descendants of $W_\pm^{(\ell)}(z)$ and level 1 Virasoro descendants of
$U_\pm(z)$, $X_\pm^{(\ell)}(z)$ and $Y_\pm^{(\ell)}(z)$ enter with
coefficient zero in the above OPEs.  The non-trivial thing to check is
that there are no new Virasoro primaries of dimension ${6\over5}$ (for
the $V_\pm J$ OPEs) or ${13\over5}$ (for the $V_\pm H_\ell$ OPEs)
contributing single pole terms in \VIxv.

Following {}from the properties of the screening charge, it is not hard
to see that a BRST cohomology analysis of the type carried out in
ref.~[\CLT] can be performed starting {}from the \hh\ FSCA.
In this way we can, in principle, construct a sequence of unitary
representations of the \hh\ FSCA at special values of the background
charge.  We will return to this point in the next section.

\chapter{Relation between the various $K=8$ FSCAs}

We will now construct the FSCA based on the $\ff2$ (or $\pp2$) \tci\
operator algebras.  In other words, we simply repeat the steps
carried out in Sects.~5 and 6 using the $\ff2$ OPEs instead of
those of the $\ff1$ operator product algebra.

It turns out that $J^{(8)}$ has the identical form as $J$ defined
in \Viii.  However, because $J^{(8)}$ is constructed using
the $\ff2$ model OPEs, we find that the $\hs1$ and $\hs2$ currents
do not appear in the $J^{(8)}J^{(8)}$ OPE.  In fact, by the
row map ${\cal R}$ \IVx\ which relates the $\ff1$ and $\ff2$ OPEs,
the \tci\ fields which entered into the definition of the $H_\ell$
currents \Viv\ decouple {}from the $\ff2$ operator product algebra.
In other words, $J^{(8)}J^{(8)}$ closes only on the identity and
$J^{(8)}$.  The background charge is turned on in the \tj8 theory
as in Sect.~6; the result for $J^{(8)}$ is the same expression \VIvi.
{}From \VIx, decoupling $\hs1$ and $\hs2$ and letting $s^2\rightarrow
{2\over3}{x\over\sqrt r}$ (which is the action of the ${\cal R}$
map), we find the $\ba J^{(8)}J^{(8)}J^{(8)}\fa$ structure
constant to be
 $$
  \lambda_8^2(c)={{32}\over{45}}\left({{27}\over{5c}}-1\right)
  {{x^2}\over r},
  \eqn\VIIi
 $$
in agreement with \Ivi\ for $K=8$, the result found in ref.~[\agt].
The screening current for the \tj8 algebra are the same as those
found for the \hh\ algebra \VIxa-\VIxii, and the screening charge
commutes with $J^{(8)}$ by virtue of a $V_\pm J$ OPE identical to
that in \VIxv.

In summary, we have constructed the currents, their structure constants
and the screening charges for both $K=8$ FSCAs.  This was done for
the general \tj{K} FSCAs in ref.~[\agt].

It is well-known
that the existence of screening charges is very important.  For
example, the construction of the characters of the minimal series
in the Feigin-Fuchs approach relies on the screening current
of the Virasoro algebra [\Felder,\Rocha].  The screening charges of
the \tj K algebra were used in a similar fashion to construct the
characters of its representations [\Kastor,\CLT].  The physical
picture behind this construction can be described as follows.  In
general, the (true) Fock space of our model is generated by the repeated
operations of the negative modes of the currents (the creation operators)
on the primary states.  In the absence of background charge,
this is equivalent to the Fock space generated by the repeated operations
of the negative modes of the PF currents and the boson field on the
primary states.  We will refer to this latter Fock space as the
PF$\otimes\{\varphi,\as0\}$ Fock space.  As we turn on the background
charge $\as0$ of the boson, the PF$\otimes\{\varphi,\as0\}$ Fock
space contains states that are absent {}from the true Fock space;
furthermore, some of the states in the true Fock space become null
(\ie\ descendant {\it and\/} primary) and must be removed.  Thus, if
we wish to start {}from the PF$\otimes\{\varphi,\as0\}$ Fock space,
both these sets of spurious states must be removed to obtain the
correct Fock space.  Because they commute with the currents of
the algebra (by construction), the screening charges are the appropriate
tools to perform this surgery.

To be more precise, we can construct a BRST operator $Q$ {}from the
screening currents of the \tj{K} FSCAs [\CLT].  For a set of discrete
values of the background charge $\alpha_0$ of the boson, we can
construct the characters of the representations of the FSCA via
the BRST cohomology.  These characters turn out to be precisely
the branching functions of the $SU(2)_K\otimes SU(2)_L/SU(2)_{K+L}$,
or simply \kl KL, coset theory, where the central charge is given by
 $$
 c={2(K-1)\over K+2}+1-12\alpha_0^2
  ={3K\over K+2}+{3L\over L+2}-{3(K+L)\over K+L+2}~.
 \eqn\VIIii
 $$
Now let us turn our attention to the other algebra we have constructed
for the $K=8$ case, namely the \hh\ FSCA.  Since this algebra has
the same screening currents as the \tj8 FSCA, we expect to obtain
the same \kl 8L coset theories as
representations, \ie\ their branching functions are also representations
of the \hh\ FSCA for the central charges given in \VIIii.  However, there
are a couple of important differences between the representation
theories of these two $K=8$ FSCAs.

First, the size of the branching function representations (reflecting
the field content of the representations) of these two FSCAs are different.
In general, the (holomorphic part of the) primary fields of the
extended conformal algebra have the form
 $$
 V_{j,p}(z)=\phi^j_j(z){\rm e}^{ip\varphi(z)}~,
 \eqn\VIIiii
 $$
where $\phi^j_j$ is a PF field, and the momentum $p$ of the boson belongs
to a well-defined set of discrete momenta [\Kastor,\CLT], whose precise
values do not concern us here.  As we have seen in earlier sections, some
of the PF fields decouple {}from the current $J^{(8)}(z)$.  This means that
some of the branching functions that are present in the coset theory are
actually missing {}from the representations
of the \tj8 FSCA.  To be specific,
the $\phi^2_2$ PF field (which corresponds to $\B{12}$ and $\B{21}$ in the
\tci\ notation) decouples {}from $J^{(8)}$.  This means that the character
(and the associated branching functions) corresponding to the primary
fields $V_{2,p}$ are not representations of the \tj8 FSCA.  To generate
them we must use the \hh\ FSCA.  This is true generically for other
values of $K$ besides $K=8$.  In particular, for $K>4$, we expect the
existence of FSCAs other than the simplest \tj{K} FSCAs constructed in
ref.~[\agt].

The second point concerns the range of central charges for which
we can find unitary coset representations of two FSCAs corresponding
to the same $K$.  Since $J^{(K)}$, when written in terms of PF fields,
contains the $\phi^1_0$ field which satisfies the fusion rule [see \IIiv]
 $$
 \phi^1_0\times\phi^1_0\sim1+\phi^1_0+\phi^2_0~,
 \eqn\VIIiv
 $$
we expect generically that the OPE of $J^{(K)}$ with itself will
generate a new current $H$ involving the dimension $6/(K+2)$
$\phi^2_0$ field.  In fact, the current $H$ can be shown [\agt] to
involve level 2 PF descendants of $\phi^2_0$, and so in general has
conformal dimension $2+6/(K+2)$.  Note that for $K=8$, this is just
the value, ${13\over5}$, that we found above for the dimension of
the $H_\ell$ currents of the \hh\ algebra.  On the other hand, the
\kl KL coset models, which should
form representations of the FSCA algebras, correspond for $L=1$ to
the minimal unitary series, with central charge $c_{L=1}=
1-6/(K+2)(K+3)$.  Now, the current $J^{(K)}(z)$ operating on the
identity generates a state with conformal dimension $(K+4)/(K+2)$.
Since the unitary minimal model always contains the primary field
$\Phi_{3,1}(z)$ with conformal dimension $(K+4)/(K+2)$, this field is
identified with the current $J^{(K)}$ in the \kl K1 coset models.  The
$\Phi_{3,1}\Phi_{3,1}$ OPE is (for $K\geq4$) [\BPZ]
 $$
 \Phi_{3,1}\Phi_{3,1}\sim1+\lambda\Phi_{3,1}+\mu\Phi_{5,1}~,
 \eqn\VIIv
 $$
where $\Phi_{5,1}$ is the dimension $4+6/(K+2)$ primary field, and
$\lambda$ and $\mu$ are structure constants that have to be
determined by associativity. Since the dimension of $\Phi_{5,1}$
is larger than that of the $H_\ell$ currents by two units and there is
no other field with the dimension of the $H_\ell$ in this minimal
model, an algebra involving the $H_\ell$ cannot be represented
by this minimal model.

The \tj{K} algebras avoid this problem, of
course, by decoupling the $H$ currents, as we have seen explicitly
above in the $K=8$ case.  This corresponds to an associativity solution
in which $\mu=0$ in \VIIv.  This problem appears to be present, though,
if we try to apply the \hh\ algebra to the \kl 81
coset model, which is identified as the eighth member of the unitary
minimal
series with central charge $c_{L=1}={52\over55}$ and which does not
have a primary field of dimension ${13\over5}$.  It turns out that in
this case the problem is resolved again by the decoupling of the $\hs\ell$
currents {}from the \hh\ FSCA.  Indeed, precisely at $c=c_{L=1}$
the \hh\ algebra structure constant $\Omega(c)$ vanishes \Iix, so that
the $\hs\ell$ decouple in the $J(z)J(w)$ OPE.  More to the point, the
$\Upsilon(c)$ structure constant \Ix\ diverges at this special value of
the central charge, showing that the
$\hs\ell$ currents must decouple {}from
the algebra as a whole. That is, to avoid the divergence we must
renormalize $\hs\ell$ resulting in their having zero norm and hence
their being null states in this representation.

Note that the resulting reduced $\{T,J\}$ algebra at
$c=c_{L=1}$ is still different {}from the \tj8 FSCA; in particular they have
different structure constants for coupling three $J$ currents [corresponding
to the $\lambda$ structure constant in \VIIv].  Since $\lambda$ for
the reduced $\{T,J\}$ algebra is different {}from the value it takes in
the \tj8 algebra, by associativity of the $\Phi_{3,1}$ four-point function,
$\mu$ in eq.~\VIIv\ must be different for the two algebras also.
In particular, $\mu$ will be non-zero for the reduced $\{T,J\}$
algebra, and thus the dimension $23\over5$ $\Phi_{5,1}$ field will
couple to these currents.  This new current actually enters with
multiplicity two, and will be denoted $\hs1'(z)$ and $\hs2'(z)$.
For $c>{52\over55}$ the $\hs\ell'$ are level-2 current algebra descendants
of the $\hs\ell$, but at $c={52\over55}$ the $\hs\ell$ are null so that the
$\hs\ell'$ are the FSCA primary currents.

Using the \tci$\otimes\{\varphi,\as0\}$ representation
for the $K=8$ FSCA developed in this paper,
the form for $\hs\ell'$ could be constructed and the
structure constants analogous to \Iix-\Ix\ for this $\{T,J,\hs\ell'\}$
FSCA could, in principle, be calculated.  In practice, however,
this would require a significant expansion of the calculation
of Virasoro descendants given in Appendix C.
Fortunately, we can use the methods
of ref. [\DoFa] to calculate directly in the $c={52\over55}$ unitary model.
An additional advantage is that we can compare the $\ba JJJ\fa$ structure
constant found {}from either method and check for consistency of our
whole picture.

To start with, the fusion rules for the unitary model force us to
consider, in addition to the fields $\p1\equiv1$, $\p3$ and $\p5$, the
fields $\p7$ and $\p9$. These additional fields do not destroy the basic
$K=8$ FSCA, because modulo 1 the $\p7$ and $\p9$ fields have the
same dimension as $\p3$ and $\p1$, respectively.
We make the following definitions
 $$\eqalign{
  1&\sim\Phi_{1,1}\qquad\qquad\Delta=0~,\cr
  J&\sim\Phi_{3,1}\qquad\qquad\Delta={6\over5}~,\cr
  \hs1',~\hs2'&\sim\Phi_{5,1}\qquad\qquad\Delta={23\over5}~,\cr
  J'&\sim\Phi_{7,1}\qquad\qquad\Delta={51\over5}~,\cr
  I'&\sim\Phi_{9,1}\qquad\qquad\Delta=18~,\cr}
  \eqn\VIIvi
 $$
where the conformal dimensions, $\Delta$, of these fields
appear in the right hand column.
Therefore, we see that we can interpret $J'$ as a
level-9 current algebra descendant of $J$,
and $I'$ as a level-18 current algebra descendant of the identity.

Calculating directly in the $c={52\over55}$ unitary model we can
construct the $\{T,J,\hs\ell'\}$ FSCA for this central charge,
analogous to \Iviii-\Ix, and we find explicitly that
 $$\eqalign{
  J\ J&=1+rx\f333J+\srx\f335\hs1'+\srx\f335\hs2'\cr
  J\ \hs1'&=\srx\f335\left[J+\fr{\f357}{\f335}J'\right]+\f355\hs2'\cr
  J\ \hs2'&=\srx\f335\left[J-\fr{\f357}{\f335}J'\right]+\f355\hs1'\cr
  \hs1'\ \hs1'&=\left[1+\f559I'\right]+\srx\f555\hs1'\cr
  \hs2'\ \hs2'&=\left[1-\f559I'\right]+\srx\f555\hs2'\cr
  \hs1'\ \hs2'&=\f355\left[J+\fr{\f5\fb7}{\f355}J'\right]\cr}
 \eqn\VIIvii
 $$
where the $f_{ijk}$ obey the structure constant symmetry properties
\IIIiii\ and \IIIv, and are given by
 $$\eqalign{
  \f333&=\fr{2\cdot7}3\sf2{13}\qquad
  \f335=\fr{5^2}{13}\sf{2\cdot5\cdot7}{17\cdot23}\cr
  \f355&=\fr{3\cdot23}{7\cdot17}\sqrt{2\cdot13}\cr
  \f555&=-\fr{3^4\cdot19\cdot37\cdot47}{5^2\cdot7\cdot13}
    \sf2{5\cdot7\cdot17\cdot23}~.\cr}
 \eqn\VIIviii
 $$
Now we can make the non-trivial check that {}from eq.~\VIIviii\ and
eq.~\Iix\ we have
 $$
  s^2\Lambda(c={52\over55})=rx\f333,
  \eqn\VIIix
 $$
so that the $\{T,J,\hs\ell'\}$ FSCA is
indeed consistent with the \hh\ one.

In the associativity constraints of the $c={52\over55}$ unitary model,
the fields $J'$ and $I'$ are not automatically included in the current
blocks of $J$ and 1, respectively, as
they are in the \tci$\otimes\{\varphi,\as0\}$ representation. Therefore,
they must be included as separate fields and their associativity
constraints checked also. We find the unique associative solution
 $$\eqalign{
  I'\ I'&=1\cr
  I'\ J&=+\f379J'\cr
  I'\ J'&=-\f379J\cr
  I'\ \hs1'&=+\f559\hs1'\cr
  I'\ \hs2'&=-\f559\hs2'\cr
  J\ J'&=\f379I'+rx\f377J'-\srx\f357\hs1'+\srx\f357\hs2'\cr
  J'\ J'&=1+rx\f377J+\srx\f577\hs1'+\srx\f577\hs2'\cr
  J'\ \hs1'&=-\srx\f357J+\srx\f577J'+\f5\fb7\hs2'\cr
  J'\ \hs2'&=+\srx\f357J+\srx\f577J'-\f5\fb7\hs1'\cr}
 \eqn\VIIx
 $$
where the new structure constants are given by
 $$\eqalign{
  \f357&=-i\fr1{5^3}\sf{3\cdot17\cdot19\cdot37\cdot47}{2\cdot7\cdot13}\cr
  \f377&=-\fr{17\cdot19\cdot29}{3\cdot13\cdot23}\sf2{13}\cr
  \f379&=i\sf{2\cdot7\cdot23\cdot29\cdot59\cdot79}{3\cdot13\cdot17\cdot37
    \cdot47\cdot67}\cr
  \f5\fb7&=-i\fr{3\cdot19\cdot29}{5^5\cdot7^2}\sf{2\cdot3\cdot19\cdot23
    \cdot37\cdot47}5\cr
  \f559&=\fr{23}{5^5}\sf{17\cdot19\cdot29\cdot59\cdot79}
    {2\cdot5\cdot7\cdot67}\cr
  \f577&=\fr{19^2\cdot29^2\cdot59\cdot79}{13\cdot5^8}
    \sf{17}{2\cdot5\cdot7\cdot23}~.\cr}
 \eqn\VIIxi
 $$
Note that the $\f5\fb7$ structure constant has one of its indices
barred to remind the reader that this structure constant
actually couples the distinct fields $\hs1'$ and $\hs2'$.  Thus,
by the structure constant symmetry \IIIv, $\f5\fb7=-\f\fb57$;
this distinction between $\hs1'$ and $\hs2'$ does not have to be
made in any other structure constant since the relevant ones are
all real.

Since the
$\Omega$ and $\Upsilon$ structure constants of the \hh\ FSCA are both
finite at other values of the central charge, we should
not encounter a similar decoupling of the $\hs1$ and $\hs2$
in any other representation of the $K=8$
\hh\ FSCA.  Indeed, it is easy to check that the $K=8$, $L\geq2$ coset
models all have primary fields with conformal dimension $13\over5$.

The Feigin-Fuchs framework gives a plausible explanation of this special
behavior of the FSCA representations for low-$L$ cosets.  The BRST
cohomology argument is based on turning on
the background charge of the boson while
leaving the PF part of the theory untouched.  As the background charge
increases, the effective central charge of the boson decreases.  The BRST
cohomology can be viewed as a reduction of the size of the Fock space to
suit the reduced central charge.  However, conservatively, we may want to
avoid having to reduce the central charge of the boson to less than zero,
since in that case we will have to reduce the size of the Fock space of the
PF model itself.  For the $K=8$ case, \ie\ for the \kl 8L
coset models, we see that the boson has zero central charge
precisely when $L=2$, and it is easy to check that the primary field
with conformal dimension $13\over5$  is still present in the $L=2$ coset
model.  However, the $L=1$ model (the minimal model), whose boson has
effective central charge $c_\varphi=-{5\over11}$,
no longer has a dimension ${13\over5}$ field.

\chapter{Application to the $K=8$ Fractional Superstring}

In this section, we turn to the application of the $K=8$ FSCAs
to fractional superstrings.  Briefly, the $K=8$ fractional
superstring [\ArT] propagates in four-dimensional Minkowski
space-time, and has a supersymmetric particle spectrum.
On the (two-dimensional) string world-sheet, the fractional
superstring is built {}from four copies of the
\zk8 theories and four coordinate
bosons (denoted $X^\mu$).  Thus, the CFT underlying this string
is a four-fold tensor product of the $c={12\over5}$ theory
discussed in Sect.~5, with a Minkowski metric.  In particular,
the free boson fields
(with no background charge) of each of the four $c={12\over5}$ theories,
which we called $\varphi$, are to be identified with the coordinate
bosons $X^\mu$ ($\mu=0,1,2,3$) of the string theory, whose radii are
infinite. The
connection between the world-sheet and space-time field content
of the $K=8$ fractional superstring is made by way of the $K=8$ FSCA.
Basically, the FSCA currents generate the physical state conditions
on the string Fock space.  This means that the non-negative modes
of the currents annihilate the physical states of the fractional
superstring.  More detailed discussions of fractional
superstring theories can be found in refs.~[\ArT,\FSS].

We have constructed in this paper two inequivalent FSCAs at $K=8$.
It is natural to ask which is the correct one to describe the
$K=8$ fractional superstring.  We will argue below, by an
examination of the physical state conditions for the massless
fermion states of the fractional superstring, that the
\hh\ FSCA is the relevant algebra.

In the $K=8$ fractional superstring the space-time
fermion states come {}from the $\phi^2_{\pm2}$ PF fields on the
world-sheet [\ArT], which are the $\B{12}$ and $\B{21}$ fields in our
\tci\ notation.  To be more precise,
the massless fermion state in four space-time dimensions is
given by
 $$
  |\Psi\fa=\left(\prod^3_{\mu=0}\sigma^\mu(z)\right)
  {\rm e}^{ip\cdot X(z)}|0\fa,
  \eqn\VIIIi
 $$
where $\sigma^\mu(z)$ stands for either $\beta_{12}^\mu(z)$ or
$\beta_{21}^\mu(z)$.  The $\mu$ index reflects the fact that we are
tensoring together four copies of the \zk8 theory as well as the
coordinate bosons $X^\mu(z)$ to obtain a Minkowski space-time
interpretation. Thus $p^\mu$ is the Minkowski space-time momentum.
The only non-trivial physical state conditions for the state \VIIIi\
(\ie, the only non-negative modes of the FSCA currents that do
not identically annihilate that state) are $L_0$ and $J_0$.
Here $L_0$ is the zero mode of the total energy-momentum tensor;
the effect of its physical state condition is simply to show
that \VIIIi\ is massless.  $J_0$ refers to either the zero mode
of the (total) $J^{(8)}$ current of the \tj8 FSCA, {\it or\/}
the zero mode of the (total) $J$ current of the \hh\ FSCA.

Since, by the constructions of Sects.~5 and 7, both $J$ currents
have the same form, namely,
 $$
 J(z)={1\over\sqrt2}\sum_{\mu=0}^3\left(\B{22}^\mu(z)\partial X_\mu(z)
 +(\B{33})^{\mu}_{{\phantom \mu}\mu}(z)\right)
 \eqn\VIIIii
 $$
(note that the total current for the tensor product theory is the
sum of the currents for each factor), it might seem that there
will be no difference in their action on the state \VIIIi.
However, in the \tj8 FSCA we learned that the $\B{12}$ and $\B{21}$
fields actually decouple {}from the fractional supercurrent $J^{(8)}$.
This means that there is no $J_0$ physical state condition on the
state \VIIIi.  Since the Dirac equation for the massless fermion
state $|\Psi\fa$ should come {}from the $J_0$ physical state condition
(and there is no other condition it could come {}from), it is clear
that the \tj8 FSCA cannot be the appropriate worldsheet symmetry
for the $K=8$ fractional superstring.

Let us see the consequences of the $J_0$ physical state
condition
 $$
 J_0|\Psi\fa~=~0~,
 \eqn\VIIIiii
 $$
when $J$ is the \hh\ FSCA current.
In this case, by the $\ff1$ \tci\ operator product algebra, the
$\B{12}$ and $\B{21}$ fields couple to the $\B{22}$ field.  Following
{}from \IVxiii\ we have the OPEs
 $$\eqalign{
  \B{22}(z)\B{12}(w)=&+i(z-w)^{-{1\over5}}\B{21}(w)
   -is(z-w)^{2\over5}\B{23}(w),\cr
  \B{22}(z)\B{21}(w)=&-i(z-w)^{-{1\over5}}\B{12}(w)
   +is(z-w)^{2\over5}\B{32}(w).\cr}
  \eqn\VIIIiv
 $$
Let us consider the moding of $\B{22}$.  Define
 $$
  |\B{ij}\fa={\rm lim}_{z\rightarrow0}\B{ij}(z)|0\fa,
  \eqn\VIIIv
 $$
then, following {}from the OPEs \VIIIiv\ and the dimensions of the
$\beta_{ij}$ fields, we find
 $$\eqalign{
  (\B{22})_0|\B{12}\fa=&+i|\B{21}\fa,\cr
  (\B{22})_0|\B{21}\fa=&-i|\B{12}\fa,\cr
  (\B{22})_{-{3\over5}}|\B{12}\fa=&-is|\B{23}\fa,\cr
  (\B{22})_{-{3\over5}}|\B{21}\fa=&+is|\B{32}\fa.\cr}
  \eqn\VIIIiv
 $$
Thus $(\B{22})_0(\B{22})_0|\sigma\fa=|\sigma\fa$ for
$|\sigma\fa=|\B{12}\fa$ or $|\B{21}\fa$.  If we now add back in the
space-time index $\mu$, and demand that the $\B{22}$ fields corresponding
to different space-time dimensions anticommute (which we can
always do by the inclusion of appropriate Klein factors),
we obtain the following anticommutation relations
 $$
  \{(\B{22})^\mu_0,(\B{22})^\nu_0\}|\Psi\fa=g^{\mu\nu}
  |\Psi\fa~.
  \eqn\VIIIv
 $$
This is just the Clifford algebra acting on $|\Psi\fa$.  Thus
$|\Psi\fa$ lies in a spinor representation of the
four-dimensional Lorentz algebra, so that we can write
 $$
 |\Psi\fa~=~|\alpha,p\fa u_\alpha(p)~.
 \eqn\VIIIva
 $$
Here $u_\alpha(p)$ is the Dirac spinor wave-function of the
massless state.

Using the explicit form for $J(z)$ given in eq.~\VIIIii\ we
can easily derive the equation of motion satisfied by the
$u_\alpha(p)$ spinor wave-function {}from the $J_0$ physical state
condition.  Specifically, by virtue of \VIIIv, the $\B{22}$ zero
modes can be identified with
Dirac gamma matrices when acting on $|\Psi\fa$:
  $$
  (\B{22})^\mu_0~=~{i\gamma^\mu\over\sqrt2}~.
  \eqn\VIIIvi
  $$
Since the $\B{33}$ term in $J_0$ automatically annihilates
$|\Psi\fa$ ({}from dimensional considerations), it follows
{}from the physical state condition \VIIIiii\ that
 $$
  J_0|\Psi\fa={1\over\sqrt2}\sum^3_{\mu=0}(\B{22})^\mu_0p_\mu|\Psi\fa
  ={i\over2} {p\hskip-6pt /}|\Psi\fa=0~,
  \eqn\VIIIvii
 $$
and thus ${p\hskip-6pt /}u=0$.
This is the Dirac equation for a massless (space-time) fermion.
This leads us to conclude that the \hh\ FSCA is the
correct worldsheet symmetry algebra for the $K=8$ fractional superstring
theory. Of course, the demonstration of the consistency of the $K=8$
fractional superstring as a whole involves a highly non-trivial
analysis that remains to be carried out.

\noindent{\bf Acknowledgements}\ \ It is our pleasure to thank
our colleagues at Cornell, and in particular E.~Angelos and A.~LeClair
for useful discussions.  This work was
supported in part by the National Science Foundation.

\Appendix{A}

In this appendix we give a very brief review of the Feigin-Fuchs technique
as applied by Dotsenko and Fateev [\DoFa] to the minimal models.
Actually we will focus only on the unitary series.  We then tabulate
all the fusion matrices for the TCI model.

The basic idea is to represent a unitary model with
$c=1-{6\over p(p+1)}$,
by a single free boson with background charge. The most important
information necessary to construct associative
algebras is the transformation
properties of the conformal blocks for four-point functions. Within the
bounds of the unitary series this can be done by solving the
differential equations that the correlation functions satisfy. As these
differential equations are constructed by using the null states in the
models, for large $p$ the differential equations will become practically
intractable. It turns out that since we are interested in only the second
member of the unitary series, with $p=4$, the differential equations
technique {\it can} be applied---in fact
that is the method used by Qiu in ref.~[\Qiu].  But even at
this low a level the methods of Dotsenko and Fateev [\DoFa] are
clearly easier.  They map the minimal model primary fields onto
exponentials of a free boson with background charge, thereby expressing
correlation functions as multiple integrals.  The normalizations for
these integrals and some of their transformation properties under
fusions were first computed by Dotsenko and Fateev; a general formula
for the transformation matrices under fusion is derived in ref.~[\agt].

Consider a general four-point function of primary fields in some level $p$
unitary model,
 $$
  G = \ba\phi_i\phi_j\phi_k\phi_l\fa.
  \eqn\Ai
 $$
(Usually, minimal model primaries are written $\Phi_{n,m}$, but we
are merely being symbolic as we do not want to hide the ideas behind too
much notational baggage.)
Taking the normalization $\phi_i\phi_j=\delta_{i,j}$ the function $G$ can
be expanded in the following two ways,
 $$\eqalign{
  G_{i\rightarrow j,\ k\rightarrow l} &= \sum_m c_{ijm}c_{klm}\ff_m,\cr
  G_{j\rightarrow k,\ l\rightarrow i} &= \sum_n c_{jkn}c_{lin}\ff'_n,\cr}
  \eqn\Aii
 $$
where $\ff_m$ and $\ff'_n$ are the conformal blocks, properly normalized,
and the $c_{ijk}$ are the structure constants appearing in the primary
field OPEs:
 $$
 \phi_i\phi_j=\sum_k c_{ijk}\phi_k~.
 \eqn\Aiia
 $$
Now using the normalization integrals of ref.~[\DoFa] and the
transformation matrices of ref.~[\agt], it is straightforward to
construct the ``fusion matrix'' relating the two different sets
of conformal blocks in \Aii.  We will always denote the fusion
matrices by $\alpha$.  They satisfy the following matrix equation
involving the structure constants:
 $$
  \sum_m\as{{n,m}}c_{ijm}c_{klm}=c_{jkn}c_{lin}.
  \eqn\Aiii
 $$
We refer to this equation as the associativity condition for the
$\ba\phi_i\phi_j\phi_k\phi_l\fa$ four-point function.
For any algebra (not just the minimal models), given
the complete set of $\alpha$ matrices, all associative algebras
can be constructed.

We now present an exhaustive list of the $\alpha$ matrices for the TCI
model. For ease of notation we abbreviate the fields in the
correlation functions by $i$ and $\h j$, where
 $$
  i=\fpi1i,\qquad \h j=\fpi2j.
  \eqn\Aiv
 $$
One appealing feature of this notation is that the field 1 is actually the
identity. Now the following table is a complete list of the $\alpha$
matrices for all four-point functions with only one conformal block.
 $$\vbox{\offinterlineskip
  \halign{ # & \hfil# & \vrule# & \quad # & \hfil# & \vrule# & \quad
  # & \hfil# & \vrule# & \quad # & \hfil# \cr
  4-point & $\alpha$ && 4-point & $\alpha$ && 4-point & $\alpha$ &&
  4-point & $\alpha$ \cr
  \omit&\omit&height3pt&\omit&\omit&&\omit&\omit&&\omit&\omit\cr
  \noalign{\hrule} \cr
  \omit&\omit&height2pt&\omit&\omit&&\omit&\omit&&\omit&\omit\cr
  $\ba2224\fa$ & $1$ && $\ba22\h1\h1\fa$ & ${1\over2}$ && $\ba22\h1\h2\fa$
  & $1$ &&
  $\ba24\h2\h2\fa$ & ${1\over2}$ \cr
  \omit&\omit&height2pt&\omit&\omit&&\omit&\omit&&\omit&\omit\cr
  $\ba2244\fa$ & ${3\over7}$ && $\ba2\h12\h1\fa$ & $-1$ && $\ba2\h12\h2\fa$
  & $1$ &&
  $\ba2\h24\h2\fa$ & $-1$ \cr
  \omit&\omit&height2pt&\omit&\omit&&\omit&\omit&&\omit&\omit\cr
  $\ba2424\fa$ & $1$ && $\ba23\h1\h1\fa$ & $1$ && $\ba23\h1\h2\fa$
  & ${3\over2}$ &&
  $\ba34\h2\h2\fa$ & ${1\over{12}}$ \cr
  \omit&\omit&height2pt&\omit&\omit&&\omit&\omit&&\omit&\omit\cr
  $\ba2334\fa$ & $-1$ && $\ba2\h13\h1\fa$ & $1$ && $\ba32\h1\h2\fa$
  & $-{1\over2}$ &&
  $\ba3\h24\h2\fa$ & $1$ \cr
  \omit&\omit&height2pt&\omit&\omit&&\omit&\omit&&\omit&\omit\cr
  $\ba2343\fa$ & $1$ && $\ba33\h1\h1\fa$ & ${3\over4}$ && $\ba2\h13\h2\fa$
  & $3$ &&
  $\ba44\h2\h2\fa$ & ${1\over{56}}$ \cr
  \omit&\omit&height2pt&\omit&\omit&&\omit&\omit&&\omit&\omit\cr
  $\ba3344\fa$ & ${3\over7}$ && $\ba3\h13\h1\fa$ & $1$ && $\ba24\h1\h2\fa$
  & ${7\over6}$ &&
  $\ba4\h24\h2\fa$ & $-1$ \cr
  \omit&\omit&height2pt&\omit&\omit&&\omit&\omit&&\omit&\omit\cr
  $\ba3434\fa$ & $1$ && $\ba44\h1\h1\fa$ & ${7\over8}$ && $\ba42\h1\h2\fa$
  & $-{1\over6}$ &&
  ${}$ & ${}$ \cr
  \omit&\omit&height2pt&\omit&\omit&&\omit&\omit&&\omit&\omit\cr
  $\ba4444\fa$ & $1$ && $\ba4\h14\h1\fa$
  & $-1$ && $\ba2\h14\h2\fa$ & $7$ &&
  ${}$ & ${}$ \cr
  \omit&\omit&\omit&\omit&\omit&height5pt&\omit&\omit&&\omit&\omit\cr
  ${}$&${}$ &\omit& ${}$ & ${}$ && $\ba33\h1\h2\fa$ & ${1\over2}$ && ${}$
  & ${}$ \cr
  \omit&\omit&\omit&\omit&\omit&height2pt&\omit&\omit&&\omit&\omit\cr
  ${}$&${}$ &\omit& ${}$ & ${}$ &&
  $\ba3\h13\h2\fa$ & $-1$ && ${}$ & ${}$ \cr
  \omit&\omit&\omit&\omit&\omit&height5pt&\omit&\omit&&\omit&\omit\cr
  ${}$&${}$ &\omit& ${}$ & ${}$ && $\ba34\h1\h2\fa$ & ${7\over4}$ && ${}$
  & ${}$ \cr
  \omit&\omit&\omit&\omit&\omit&height2pt&\omit&\omit&&\omit&\omit\cr
  ${}$&${}$ &\omit& ${}$ & ${}$ && $\ba43\h1\h2\fa$ & $-{1\over4}$ && ${}$
  & ${}$ \cr
  \omit&\omit&\omit&\omit&\omit&height2pt&\omit&\omit&&\omit&\omit\cr
  ${}$&${}$ &\omit& ${}$ & ${}$ && $\ba3\h14\h2\fa$
  & $-7$ && ${}$ & ${}$ \cr}}
  \eqn\Av
 $$
Permuting the order of the fields in the four-point functions in \Av\ does
not lead to new associativity constraints, by virtue of the symmetries
of the structure constants \IIIiii\ and \IIIv, and by the form of the
associativity constraint \Aiii.

As an example of how to use the above table, consider the four-point
function $\ba34\h1\h2\fa$. Since $\alpha={7\over4}$ for this correlation
function we can write,
 $$
  {7\over4}c_{342}c_{\h1\h22}=c_{4\h1\h1}c_{\h23\h1}.
  \eqn\Avi
 $$
Of course, we know that a single chiral TCI model has no consistent
solution with the $\h1$ and $\h2$ fields present (see Sect.~3), but
tensoring this
four-point function with itself and then solving gives the following
constraint
 $$
  {{49}\over{16}}c_{(33)(44)(22)}c_{(\h1\h1)(\h2\h2)(22)}=
  c_{(44)(\h1\h1)(\h1\h1)}c_{(\h2\h2)(33)(\h1\h1)}.
  \eqn\Avii
 $$
Identifying the $ii$ indices with $\beta$ fields and the $\h j\h j$ indices
with either $\lambda$ or $\bar\lambda$ fields (see Sect.~4) gives an
associativity constraint on the structure constants of the \tci\ model.
In Appendix B, where a complete solution to the associativity constraints
of the \tci\ model involving the $\lambda$ and $\bar\lambda$ fields is
given, one can check that \Avii\ is indeed satisfied.

The following table summarizes all of the $\alpha$ matrices for those
four-point functions with exactly two conformal blocks.
 $$\vbox{\offinterlineskip
 \halign{&\ $\shrinker#\ $&\hfil$\shrinker#$\hfil&\vrule#\cr
 {\rm 4-point}&\alpha&&{\rm 4-point}&\alpha&&{\rm 4-point}&\alpha\cr
 \omit&\omit&height3pt&\omit&\omit&&\omit&\omit\cr
 \noalign{\hrule}\cr
 \omit&\omit&height3pt&\omit&\omit&&\omit&\omit\cr
 \ba2222\fa&\mtt{r}{{3\over2}{r\over x}}{{2\over3}x}{-r}&&
 \ba22\h2\h2\fa&\mtt{{1\over2}r}{-{3\over2}{r\over x}}{x}{3r}&&
 \ba\h1\h1\h1\h1\fa&{1\over\sqrt2}\mtt{1}{8\over7}{7\over8}{-1}\cr
  \omit&\omit&height5pt&\omit&\omit&&\omit&\omit\cr
 \ba3333\fa&\mtt{r}{{3\over2}{r\over x}}{{2\over3}x}{-r}&&
 \ba2\h22\h2\fa&\mtt{-r}{{1\over2}{r\over x}}{2x}{r}&&
 \ba\h1\h1\h2\h2\fa&{1\over\sqrt2}\mtt{1\over2}{4}{3\over4}{-6}\cr
  \omit&\omit&height5pt&\omit&\omit&&\omit&\omit\cr
 \ba2233\fa&\mtt{{2\over3}x}{r}
  {{3\over7}r}{-{9\over{14}}{r\over x}}&&
 \ba23\h2\h2\fa&\mtt{{3\over4}{r\over x}}{-7r}{{1\over2}r}
  {{{14}\over3}x}&&
 \ba\h1\h2\h1\h2\fa&{1\over\sqrt2}\mtt{-1}{2\over3}{3\over2}{1}\cr
  \omit&\omit&height5pt&\omit&\omit&&\omit&\omit\cr
 \ba2323\fa&\mtt{-r}{{{14}\over9}x}{{9\over{14}}{r\over x}}{r}&&
 \ba2\h23\h2\fa&\mtt{r}{{3\over2}{r\over x}}{{2\over3}x}{-r}&&
 \ba\h1\h2\h2\h2\fa&{1\over\sqrt2}\mtt{1}{2}{1\over2}{-1}\cr
  \omit&\omit&height5pt&\omit&\omit&&\omit&\omit\cr
 {}&{}&&
 \ba33\h2\h2\fa&\mtt{{3\over4}r}{{9\over4}{r\over x}}{{1\over6}x}
  {-{1\over2}r}&&{}&{}\cr
  \omit&\omit&height5pt&\omit&\omit&&\omit&\omit\cr
 {}&{}&&
 \ba3\h23\h2\fa&\mtt{r}{-{9\over2}{r\over x}}{-{2\over9}x}{-r}&&
 {}&{}\cr}}
  \eqn\Aviii
 $$
A few words of explanation about this table are in order.
We will explain the conventions used in \Aviii\ by way of an
example.  For instance, the $\ba23\h2\h2\fa$ four-point function
gives rise to the associativity condition
 $$
  \mtt{{3\over4}{r\over x}}{-7r}{{1\over2}r}
  {{{14}\over3}x}\pmatrix{c_{232}c_{\h2\h22}\cr c_{234}c_{\h2\h24}\cr}
  =\pmatrix{c_{3\h2\h1}c_{\h22\h1}\cr c_{3\h2\h2}c_{\h22\h2}\cr}.
  \eqn\Aix
 $$
The pattern in eq.~\Aix\ persists for all entries in the table.
That is, the fields that are fused to, \ie\ the third index of the
$c_{ijk}$s, are placed in
increasing numerical order down the column matrix of structure constants.

The final four-point function to consider is the only one with more
than two conformal blocks, that is $\ba\h2\h2\h2\h2\fa$. It has,
in fact, four conformal blocks and implies the associativity constraint
 $$
  \ba\h2\h2\h2\h2\fa \Rightarrow {1\over\sqrt2}
  \pmatrix{ r & {r\over x} & 6{r\over x} &  56r \cr
  x & r & -6r & -56x \cr
  {1\over6}x & -{1\over6}r & -r & {{28}\over3}x \cr
  {1\over{56}}r & -{1\over{56}}{r\over x}
  & {3\over{28}}{r\over x} & -r \cr}
  \pmatrix{c^2_{\h2\h21} \cr c^2_{\h2\h22} \cr c^2_{\h2\h23} \cr
  c^2_{\h2\h24} \cr} =
  \pmatrix{c^2_{\h2\h21} \cr c^2_{\h2\h22} \cr c^2_{\h2\h23} \cr
  c^2_{\h2\h24} \cr}.
  \eqn\Ax
 $$
We see that the convention used for the table of two conformal block
four-point functions persists here also. This completes the associative
information about the TCI model needed to construct consistent algebras.

\vfill\eject

\Appendix{B}

In this appendix we present a compilation of the $\ff1$ \tci\ OPEs
involving the $\lambda$ and $\bar\lambda$ fields.  The $\lambda\lambda$
OPEs are
 $$\eqalign{
  \la{11}\la{11} & = \B{11} - {7\over8}\B{44},\cr
  \la{11}\la{12}&=-\la{12}\la{11}= i\sf34\B{13}-
  i\sf7{16}\B{42},\cr
  \la{11}\la{21}&=-\la{21}\la{11}= i\sf34\B{31}-
  i\sf7{16}\B{24},\cr
  \la{11}\la{22}&=+\la{22}\la{11}=-
  {3\over4}\B{33}+{1\over2}\B{22},\cr
  \la{12}\la{21}&=+\la{21}\la{12}=+
  {3\over4}\B{33}-{1\over2}\B{22},\cr
  \la{12}\la{12}&=\B{11}+\sf{rx}6\B{13}
  +\sf{7rx}8\B{42}+{1\over8}\B{44},\cr
  \la{21}\la{21}&=\B{11}+\sf{rx}6\B{31}
  +\sf{7rx}8\B{24}+{1\over8}\B{44},\cr
  \la{21}\la{22}&=-\la{22}\la{21}= i\sf34\B{13}+i\sf{rx}8\B{33}
  +i\sf7{16}\B{42}+i\sf{rx}2\B{22},\cr
  \la{12}\la{22}&=-\la{22}\la{12}= i\sf34\B{31}+i\sf{rx}8\B{33}
  +i\sf7{16}\B{24}+i\sf{rx}2\B{22},\cr
  \la{22}\la{22} & = \B{11} + \sf{rx}6
  (\B{13}+\B{31}) + {{rx}\over6}\B{33}\cr
  &\qquad-rx\B{22}-\sf{rx}{56}(\B{42}+\B{24})-{1\over{56}}\B{44}~,\cr}
  \eqn\Bi
 $$
According to our table of \tci\ field groupings \IVii, these OPEs can be
symbolically written
 $$
  \lambda\cdot\lambda \sim \beta^{(1)}+\beta^{(2)}.
  \eqn\Bii
 $$
Now the $\bar\lambda\bar\lambda$ OPEs are easily constructed {}from
the $\lambda\lambda$ ones by writing
 $$
  \bar\lambda\cdot\bar\lambda \sim \beta^{(1)}-\beta^{(2)}.
  \eqn\Biii
 $$
For instance
 $$\eqalign{
  &\la{11}\la{22}=-{3\over4}\B{33}+{1\over2}\B{22}\cr
  \Longrightarrow\quad&\lb{11}\lb{22}=-{3\over4}\B{33}-{1\over2}\B{22}.\cr}
  \eqn\Biv
 $$

Another set of OPEs to construct are the $\lambda\bar\lambda$ ones
which we tackle next.  The structure constants of the $\lambda\bar\lambda$
OPEs are determined by the following rule. Consider
the following `algebra' which is {\it not} associative:
 $$\eqalign{
  \fpi21\fpi21&\simdot\fpi11+\sf78\fpi14,\cr
  \fpi21\fpi22&\simdot\sf12\fpi12+\sf34\fpi13,\cr
  \fpi22\fpi22&\simdot\fpi11+\sqrt{rx}\fpi12+\sf{rx}6\fpi13+
  \sf1{56}\fpi14.\cr}
  \eqn\Bv
 $$
The $\simdot$ relation is to emphasize that \Bv\ is not a true
associative algebra, but merely a building block for one.
Taking a `direct product' (\ie\ simply multiplying the
structure constants) of the \Bv\ `algebra' with itself
yields the correct structure constants up to phases for the
$\lambda\bar\lambda$ OPEs [as well as for those in the list \Bi].
However, the determination of the proper phases can only
be gotten by explicit calculation. They are summarized in the
following table:
 $$\vbox{\offinterlineskip
 \halign{\hfil$\shrinker#$\ \hfil&&\vrule#&\hfil\ $\shrinker#$\ \hfil\cr
 \lambda\times\bar\lambda&&\lb{11}&&\lb{12}&&\lb{21}&&\lb{22}\cr
 \omit&height3pt&\omit&&\omit&&\omit&&\omit\cr
 \noalign{\hrule}\cr
 \omit&height3pt&\omit&&\omit&&\omit&&\omit\cr
 \la{11}&&+\w+\B{14}+\w-\B{41}&&-\w-\B{12}+\w+\B{43}&&
  -\w-\B{34}+\w+\B{21}&&-\w+\B{32}-\w-\B{23}\cr
 \omit&height3pt&\omit&&\omit&&\omit&&\omit\cr
 \noalign{\hrule}\cr
 \omit&height3pt&\omit&&\omit&&\omit&&\omit\cr
 \la{12}&&+\w-\B{12}-\w+\B{43}&&-\w+\B{12}+\w-\B{43}&&
  +\w+\B{32}+\w-\B{23}&&+\w-\B{32}+\w+\B{23}\cr
 \omit&height2pt&\omit&&\omit&&\omit&&\omit\cr
 {}&&{}&&-\w+\B{14}+\w-\B{41}&&{}&&+\w-\B{34}+\w+\B{21}\cr
 \omit&height3pt&\omit&&\omit&&\omit&&\omit\cr
 \noalign{\hrule}\cr
 \omit&height3pt&\omit&&\omit&&\omit&&\omit\cr
 \la{21}&&+\w-\B{34}-\w+\B{21}&&+\w+\B{32}+\w-\B{23}&&
  +\w+\B{34}-\w-\B{21}&&-\w-\B{32}-\w+\B{23}\cr
 \omit&height2pt&\omit&&\omit&&\omit&&\omit\cr
 {}&&{}&&{}&&+\w+\B{14}-\w-\B{41}&&-\w-\B{12}-\w+\B{43}\cr
 \omit&height3pt&\omit&&\omit&&\omit&&\omit\cr
 \noalign{\hrule}\cr
 \omit&height3pt&\omit&&\omit&&\omit&&\omit\cr
 \la{22}&&-\w+\B{32}-\w-\B{23}&&+\w-\B{32}+\w+\B{23}&&
  -\w-\B{32}-\w+\B{23}&&+\w+\B{32}+\w-\B{23}\cr
 \omit&height2pt&\omit&&\omit&&\omit&&\omit\cr
 {}&&{}&&+\w-\B{34}+\w+\B{21}&&-\w-\B{12}-\w+\B{43}&&
  +\w+\B{34}+\w-\B{21}\cr
 \omit&height2pt&\omit&&\omit&&\omit&&\omit\cr
 {}&&{}&&{}&&{}&&+\w+\B{12}+\w-\B{43}\cr
 \omit&height2pt&\omit&&\omit&&\omit&&\omit\cr
 {}&&{}&&{}&&{}&&+\w+\B{14}+\w-\B{41}\cr}}
 \eqn\Bviii
 $$
where $\w\pm$ are the eighth roots of unity
 $$
  \w\pm = e^{\pm i {\pi\over4}}.
  \eqn\Bvii
 $$
For example, {}from \Bv\ and \Bviii, the following OPEs can
be constructed,
 $$\eqalign{
  \la{22}\lb{12}&=\sf3{224}\w-\B{34}+\sf{3rx}4\w-\B{32}+
  \sf12\w+\B{21}+\sf{rx}{12}\w+\B{23}~,\cr
  \lb{12}\la{22}&=\sf3{224}\w+\B{34}+\sf{3rx}4\w+\B{32}+
  \sf12\w-\B{21}+\sf{rx}{12}\w-\B{23}~.\cr}
  \eqn\Bvi
 $$
The $\bar\lambda\lambda$ OPEs as well as the $\beta\lambda$
and $\beta\bar\lambda$ OPEs, can be deduced {}from the above
OPEs by the symmetries of the structure constants \IIIv.
This completes the construction of the $\ff1$ \tci\ operator algebra.

\Appendix{C}

In this appendix we use the conformal Ward identities to compute the
coefficients of the first few descendant fields appearing on the right hand
side of OPEs of fields which are themselves Virasoro descendants.
The basic principles for this derivation were first outlined
in ref.~[\BPZ], and the result (C12)-(C13) below was explicitly
calculated when $i=j$. The method outlined here is the logical
basis for any such calculation, and can be reformulated in various ways.

Let us first discuss how to compute the OPEs between descendants
of primary fields in general, given the structure constants for the
primaries themselves.  That is, we want to show how to calculate the
$\beta_{ijk}^{\vrv m\vrv n\vrv p}$ coefficients in the following OPE
 $$
  \phi_i^{\vrv m}(z)\phi_j^{\vrv n}(0) = \sum_k\sum_{\vrv p}c_{ijk}
  z^{\Delta_k-\Delta_i-\Delta_j+p-m-n}\beta_{ijk}^{\vrv m\vrv n\vrv p}
  \phi_k^{\vrv p}(0),
  \eqn\Ci
 $$
where
 $$\eqalign{
  \phi_k^{\vrv p}(0) & = (L_{-\vrv p}\phi_k)(0), \cr
  L_{-\vrv p} & = {\hat L}_{-p_1}{\hat L}_{-p_2}\cdots{\hat L}_{-p_t}, \cr
  {\hat L}_{-p_i} & = \cases{L_{-p_i},&if $p_i\ne 0$,\cr
   1,&if $p_i=0$,\cr}\cr
  p & = p_1+p_2+\cdots+p_t~, \cr}
  \eqn\Cii
 $$
and
 $$
  \beta_{ijk}^{\{0\}\{0\}\{0\}} \equiv 1~.
  \eqn\Ciii
 $$
Note that we have defined ${\hat L}_{-p}$ so that $\phi^{\vr0}(z)=\phi(z)$.

We can determine the $\beta_{ijk}^{\vrv m\vrv n\vrv p}$ by evaluating
the three-point function of descendants,
 $$
  \ba\phi_i^{\vrv m}(z_1)\phi_j^{\vrv n}(z_2)\phi_k^{\vrv p}(z_3)\fa~,
  \eqn\Civ
 $$
in two different ways.  First, take the limit as $z_1\rightarrow
z_2$ by using the OPE \Ci\ to reduce the three-point function \Civ\
to the sum over two-point functions
 $$
  c_{ijk}\sum_{\vrv q}z_{12}^{\Delta_k-\Delta_i-\Delta_j+q-m-n}
   \beta_{ijk}^{\vrv m\vrv n\vrv q}\ba\phi_k^{\vrv q}(z_2)
   \phi_k^{\vrv p}(z_3)\fa~,
  \eqn\Cv
 $$
where $z_{ij}=z_i-z_j$. The two-point functions in eq.~\Cv\ can be
calculated using the conformal Ward identities. One finds that their
coordinate dependence is $z_{23}^{-2\Delta_k-q-p}$ and their normalization
depends only on $\Delta_k$ and the central charge $c$.

On the other hand, the three-point function \Civ\ can be calculated
directly in terms of the structure constant $c_{ijk}$, again using
the conformal Ward
identities.  To be more explicit, the general descendant field
$\phi_i^{\vrv m}(z)$ can be written as the following integral
 $$
 \phi_i^{\vrv m}(z)=\oint{{\rm d}\zeta_1T(\zeta_1)\over(\zeta_1-z)^{m_1+1}}
 \cdots \oint{{\rm d}\zeta_tT(\zeta_t)\over(\zeta_t-z)^{m_t+1}}\phi_i(z)~,
 \eqn\Cvi
 $$
where the contours of integration are nested circles enclosing $z$.
Thus, the two-point functions in eq.~\Cv\ and the three-point
function \Civ\ can be expressed as the integrals
of multi-point functions involving only insertions of the energy momentum
tensor along with the primary fields.
Now, conformal invariance implies the Ward identity [\BPZ]
 $$\eqalign{
 &\ba T(\zeta_1)\cdots T(\zeta_M)\phi_{k_1}(z_1)\cdots\phi_{k_N}(z_N)\fa\cr
 &=\left\{\sum_{i=1}^N\left[{\Delta_{k_i}\over(\zeta_1-z_i)^2}
  +{1\over \zeta_1-z_i}{\partial\over\partial z_i}\right]
  +\sum_{j=2}^M\left[{2\over\zeta_{1j}^2}
  +{1\over \zeta_{1j}}{\partial\over\partial\zeta_j}\right]\right\}\cr
 &\quad\times\ba T(\zeta_2)\cdots T(\zeta_M)
  \phi_{k_1}(z_1)\cdots\phi_{k_N}(z_N)\fa\cr
 &+\sum_{j=2}^M{c\over\zeta_{1j}^4}
  \ba T(\zeta_2)\cdots T(\zeta_{j-1})T(\zeta_{j+1})\cdots T(\zeta_M)
  \phi_{k_1}(z_1)\cdots\phi_{k_N}(z_N)\fa~,\cr}
 \eqn\Cvii
 $$
where $\zeta_{ij}=\zeta_i-\zeta_j$.
Using this Ward identity repeatedly, an $n$-point function of descendant
fields can be written in terms of
the $n$-point function of the primary fields. Again by conformal
invariance, the two and three-point functions of primary fields are
 $$\eqalign{
  \ba\phi_i(z_1)\phi_j(z_2)\fa=&\delta_{ij}z_{12}^{-2\Delta_i}~,\cr
  \ba\phi_i(z_1)\phi_j(z_2)\phi_k(z_3)\fa=&
  c_{ijk}z_{12}^{-\delta_k}z_{13}^{-\delta_j}z_{23}^{-\delta_i}~,\cr}
 \eqn\Cviii
 $$
where we have introduced the useful combinations of dimensions
 $$\eqalign{
  \delta_i&=-\Delta_i+\Delta_j+\Delta_k,\cr
  \delta_j&=+\Delta_i-\Delta_j+\Delta_k,\cr
  \delta_k&=+\Delta_i+\Delta_j-\Delta_k.\cr}
  \eqn\Cix
 $$
Now if we systematically compute the three-point function \Civ\ for
all fields $\phi_k^{\vrv p}(z_3)$ using the two different methods
outlined above, then we can compare the expansion \Cv\ with the direct
calculation of eq. \Civ\ and thereby deduce the $\beta_{ijk}^
{\vrv m\vrv n\vrv p}$ coefficients.

For the purposes of this paper we only need to know the
$\beta_{ijk}^{\vrv m\vrv n\vrv p}$ for
 $$
  \vrv m,\vrv n,\vrv p \subset \{\vr0,\vr1,\vr{1,1},\vr2\}.
  \eqn\Cx
 $$
It is sufficient to calculate those with $\vrv m,\vrv n\subset
\{\vr0,\vr2\}$, since those with $\vr1$ or $\vr{1,1}$ can be simply
reached via differentiation (recall that $L_{-1}=\partial$ when acting
on a primary field). Denoting by ${\cal M}$ the matrix of inner
products of the level-$2$ Virasoro descendants of $\phi_k$:
 $$
  {\cal M} = \pmatrix{4\Delta_k(2\Delta_k+1)&6\Delta_k\cr
       6\Delta_k&4\Delta_k+{c\over2}\cr},
  \eqn\Cxiv
 $$
we present the results of these calculations in the form of the four
OPEs between the fields $\phi_i,\phi_i^{\vr2}$ and $\phi_j,\phi_j^{\vr2}$:

 $$\eqalign{
  \phi_i(z)\phi_j(0)=c_{ijk}z^{-\delta_k}&\Bigl[\phi_k(0)
  +z{{\delta_j}\over{2\Delta_k}}\phi_k^{\vr1}(0)\cr
  &\quad+z^2(a\phi_k^{\vr{1,1}}+b\phi_k^{\vr2})+\cdots\Bigr],\cr}
  \eqn\Cxv
 $$
where
 $$
  {\cal M}\pmatrix{a\cr b\cr}=\pmatrix
  {\delta_j(\delta_j+1)\cr\delta_j+\Delta_i\cr}.
  \eqn\Cxvi
 $$

 $$\eqalign{
  \phi_i^{\vr2}(z)\phi_j(0)=c_{ijk}
  z^{-\delta_k-2}&\Bigl[(\delta_k+\Delta_j)
  \phi_k(0)+z{f\over2\Delta_k}\phi_k^{\vr1}(0)\cr
  &\quad+z^2(a\phi_k^{\vr{1,1}}+b\phi_k^{\vr2})
  +\cdots\Bigr],\cr}
  \eqn\Cxvii
 $$
 $$
  f=(\delta_k+\Delta_j)(\delta_j+2)-3\delta_k~,
  \eqn\Cxviii
 $$
 $$
  {\cal M}\pmatrix{a\cr b\cr}=
  \pmatrix{f(\delta_j+3)+3\delta_j(-\delta_k+1)\cr
  (\delta_k+\Delta_j)(\delta_j+\Delta_i+4)
  -9\delta_k+4\Delta_i+{c\over2}\cr}.
  \eqn\Cxix
 $$

 $$\eqalign{
  \phi_i(z)\phi_j^{\vr2}(0)=c_{ijk}z^{-\delta_k-2}&
  \Bigl[(\delta_k+\Delta_i)\phi_k(0)+z{f\over2\Delta_k}\phi_k^{\vr1}(0)\cr
  &\quad+z^2(a\phi_k^{\vr{1,1}}+b\phi_k^{\vr2})+\cdots\Bigr],\cr}
  \eqn\Cxx
 $$
 $$
  f=(\delta_k+\Delta_i)(\delta_j-2)+3\delta_k~,
  \eqn\Cxxi
 $$
 $$
  {\cal M}\pmatrix{a\cr b\cr}=\pmatrix{f(\delta_j-1)
  +3(\delta_j\delta_k+\delta_i)\cr
  (\delta_k+\Delta_i)(\delta_j+\Delta_i-2)
  +4\Delta_j+{c\over2}\cr}.
  \eqn\Cxxii
 $$

 $$\eqalign{
  \phi_i^{\vr2}(z)\phi_j^{\vr2}(0)=c_{ijk}z^{-\delta_k-4}&
  \Bigl[g\phi_k(0)+z{f\over2\Delta_k}\phi_k^{\vr1}(0)\cr
  &\quad+z^2(a\phi_k^{\vr{1,1}}+b\phi_k^{\vr2})+\cdots\Bigr],\cr}
  \eqn\Cxxiii
 $$
 $$\eqalign{
  g={c\over2}+&11(\Delta_i+\Delta_j)-7\Delta_k
  +2\Delta_i^2+5\Delta_i\Delta_j\cr
  &+2\Delta_j^2-3(\Delta_i+\Delta_j)\Delta_k+\Delta_k^2~,\cr}
  \eqn\Cxxiv
 $$
 $$
  f=g\delta_j+3(\delta_k+2)(\Delta_j-\Delta_i)~,
  \eqn\Cxxv
 $$
 $$
  {\cal M}\pmatrix{a\cr b\cr}=\pmatrix{\left\{\matrix{\scriptstyle
  f(\delta_j+1)+6\Delta_k(\delta_k+\Delta_j)\cr
  +3(\delta_k+1)[2\Delta_k+\delta_j(\Delta_j-\Delta_i)]\cr}\right\}\cr
  \cr\left\{\matrix{\scriptstyle g(\delta_j+\Delta_i+2)
  -9(\delta_k+\Delta_i)(\delta_k+2)\cr
  \scriptstyle+(4\Delta_i+{c\over2})(\delta_k+\Delta_i)
  +(4\Delta_j+{c\over2})(\delta_k+\Delta_j)\cr}\right\}\cr}.
  \eqn\Cxxvi
 $$

\refout
\end